\documentclass[preprint]{ptephy_v2}

\preprintnumber{XXXX-XXXX} 

\usepackage{graphicx}
\usepackage{fancybox}

\usepackage{subcaption}

\title{Development and performance evaluation of a water-based liquid scintillator tracking detector with wavelength-shifting fiber readout}

\author[1]{Naoto Onda\thanks{E-mail: onda.naoto.34n@st.kyoto-u.ac.jp}}
\author[1]{Yuka Asano\thanks{Currently in industry.}}
\author[2]{Takashi Iida}
\author[1]{Tatsuya Kikawa\thanks{E-mail: kikawa.tatsuya.6e@kyoto-u.ac.jp}}
\author[1]{Tsuyoshi Nakaya}
\author[4]{Atsushi Tokiyasu}
\author[3]{Daiki Wakabayashi$^\ddagger$}

\affil[1]{Department of Physics, Kyoto University, Kyoto, Kyoto 606-8502, Japan}
\affil[2]{Institute of Pure and Applied Sciences, University of Tsukuba, Tsukuba, Ibaraki 305-8571, Japan}
\affil[3]{Department of Physics, Tohoku University, Sendai, Miyagi 980-8578, Japan}
\affil[4]{Research Center for Accelerator and Radioisotope Science, Tohoku University, Sendai, Miyagi 982-0826, Japan}

\begin{document}

\begin{abstract}
We have developed a novel tracking detector utilizing a water-based liquid scintillator (WbLS) for the accurate characterization of neutrino interactions on a water target.
In this detector, the WbLS is optically segmented into small cells by reflective separators, and the scintillation light is read out in three directions using wavelength-shifting fibers coupled to silicon photomultipliers.
We developed and optimized WbLS samples for this application and measured their light yield using cosmic-ray muons.
Subsequently, we constructed a prototype of the WbLS tracking detector and evaluated its performance with a positron beam.
The beam test demonstrated good tracking performance, although the light yield was lower than required.
The result prompted a review of the surfactant used in the WbLS and the material of the optical separators, leading to a significant improvement in light yield.
In this paper, we report on a design of the WbLS tracking detector, the development of the WbLS, the results of the beam test, and subsequent improvements to the WbLS and optical separators.

\end{abstract}

\subjectindex{Water-based liquid scintillator, Neutrino oscillation, Neutrino interaction, Wavelength-shifting fiber, Silicon photomultipliers}
\maketitle

\section{Introduction}\label{sec_intro}
Water-based liquid scintillator (WbLS) is a novel detection medium consisting of a mixture of liquid scintillator and water, typically stabilized using surfactants.
Its high water content reduces the flammability and chemical hazards associated with pure organic liquid scintillators, making it safer and more environmentally friendly.
The use of water as the primary component also improves cost-effectiveness and scalability, making WbLS a promising candidate for future large-volume detectors in neutrino and rare-event experiments.

In addition, WbLS emits both scintillation and Cherenkov light, enabling better particle identification and energy reconstruction compared to a pure water Cherenkov configuration. 
These features have made WbLS the focus of increasing research efforts in recent years~\cite{bib:yeh, bib:hans}.
Previous developments of WbLS have primarily focused on monolithic detectors consisting of a large volume of WbLS viewed by photomultiplier tubes (PMTs) on the inner surface of the detector vessel. However, these designs are not optimized for tracking individual charged particles with high granularity.

For a precise measurement of neutrino oscillations with Hyper-Kamiokande~\cite{bib:hk}, a next-generation water Cherenkov detector currently under construction, accurate characterization of neutrino interactions on a water target is essential. Although the upgraded near detector in the T2K experiment~\cite{bib:t2k} provides accurate tracking of charged particles from neutrino interactions, its target material is plastic scintillator\cite{bib:tdr, bib:sfgd, bib:sfgd_kikawa}. Since neutrino-nucleus interaction cross sections depend strongly on the target nucleus, this difference introduces a source of systematic uncertainty.
To enable precise measurements of neutrino interactions on water, we have developed a new tracking detector, in which the WbLS is optically segmented into small cells by reflective separators and the scintillation light is collected in three orthogonal directions by wavelength-shifting (WLS) fibers coupled to silicon photomultipliers (SiPMs). This structure enables three-dimensional tracking of charged particles in a water-equivalent medium with high granularity.
In this paper, we present the development and performance evaluation of the WbLS tracking detector.
The remainder of this paper is organized as follows.
Section~\ref{sec_concept} describes the design of the new detector.
Section~\ref{sec_development} presents the development of the WbLS and its initial characterization using cosmic rays.
Section~\ref{sec_beamtest} details the performance evaluation of prototype detectors using a 500 MeV positron beam at the Research Center for Accelerator and Radioisotope Science (RARiS), Tohoku University.
Section~\ref{sec_improve} summarizes the improvements to the WbLS formulation and the optical separator material based on the beam test results.
Finally, conclusions are given in Section~\ref{sec_conclusion}.

\ifpreprint
\newpage
\fi

\section{Design of new WbLS tracking detector}\label{sec_concept}

\begin{figure}[htbp]
  \begin{center}
    \includegraphics[width=0.8\linewidth]{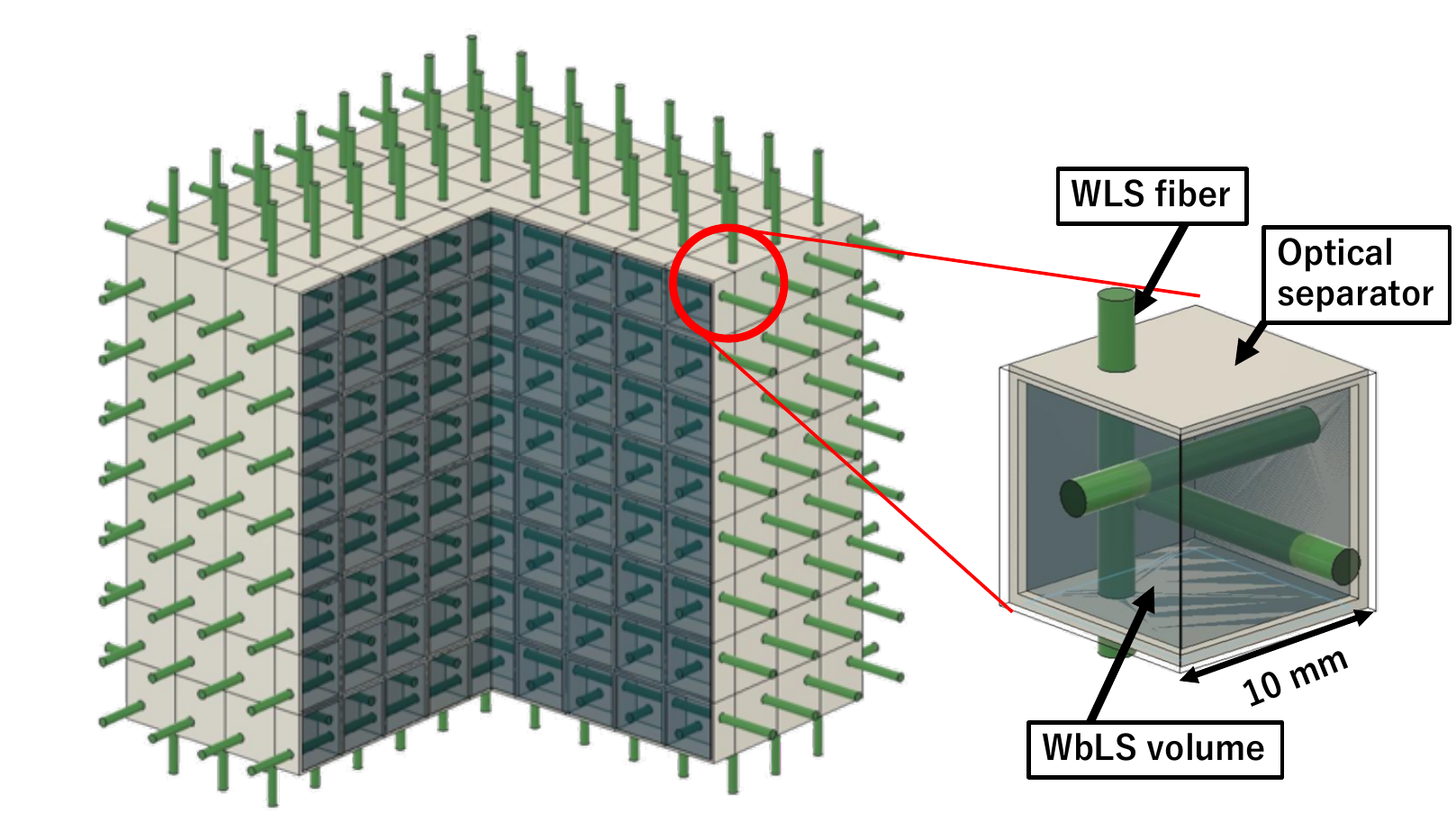}
    \caption{Schematic view of the WbLS tracking detector concept with orthogonal WLS fibers and optically segmented cells.}
    \label{wbls_concept}
  \end{center}
\end{figure}

A design of the new WbLS tracking detector is illustrated in Fig.~\ref{wbls_concept}.
The detector uses WbLS as a neutrino interaction target as well as a sensitive medium.
The WbLS is optically segmented into 1~cm $\times$ 1~cm$ \times$ 1~cm cells by reflective separators.
Holes are drilled in the separators from three orthogonal directions, and 1~mm-diameter WLS fibers are inserted along the holes.
These holes also serve as channels for filling the WbLS into individual cells.
Although the reflective separators are structurally complex, they can be manufactured by 3D printing.
The light from the fibers is detected by SiPMs coupled to the fiber ends.
This optically segmented three-dimensional structure of the detector
is similar to that of the SuperFGD\cite{bib:tdr, bib:sfgd, bib:sfgd_kikawa}, a part of the upgraded T2K near detector consisting of approximately two million plastic scintillator cubes, except that the active medium is the WbLS instead of the plastic scintillator to provide a water-equivalent target.
In this geometry, the detector enables high-resolution measurement of particles from neutrino interactions with the following capabilities as in the SuperFGD:
\begin{itemize}
\item Efficient three-dimensional reconstruction of charged particles scattered in any direction.
\item Detection of short-range, low-momentum charged hadrons, including protons down to 350~MeV/c.
\item Capability to distinguish electrons from $\gamma$-rays, enabling precise measurement of electron neutrino interactions.
\end{itemize}
This work represents the first attempt to implement WbLS in a finely segmented tracking detector with a WLS fiber readout system.

\section{Development of WbLS}\label{sec_development}

We developed the optimal WbLS for the WLS fiber readout.
The requirements to the WbLS are as follows.

\begin{description}
   \item[(1) High water fraction]\mbox{}\\
For a measurement of neutrino interactions on water, interactions occurring in non-water components of the WbLS contribute to background events.
The acceptable level of such background depends on the uncertainties associated with neutrino interactions in these additional materials.
The fluorescent materials, solvents, and surfactants incorporated into the WbLS in this study are all derived from hydrocarbon compounds.
Thus, the background contribution can be constrained by using the SuperFGD, which shares the same structural design as the WbLS tracking detector but consists entirely of hydrocarbons.
With this approach, the uncertainty in the background contribution is expected to be reduced to less than 10\%.
Taking into account that approximately 15\% of the active volume of the WbLS tracking detector is occupied by WLS fibers and reflective materials, the minimum acceptable water content in the WbLS was set at 65\%.   
   \item[(2) High light yield]\mbox{}\\
The WbLS tracking detector is designed to detect minimum ionizing particles (MIPs) with an efficiency greater than 99\% at each fiber readout.
Assuming a hit detection threshold of 1.5 photoelectrons (p.e.) and accounting for optical attenuation in the fibers, achieving this efficiency requires a light yield of at least 8.1 p.e./MeV per fiber readout channel.
To achieve a high light yield, it is essential that the emission spectrum of the WbLS matches the absorption spectrum of the WLS fibers.

   \item[(3) Long-term stability]\mbox{}\\
Neutrino experiments require stable data acquisition over several years in order to accumulate sufficient statistics.
Accordingly, the WbLS must exhibit chemical and optical stability over extended periods to ensure consistent performance throughout long-term operation.
\end{description}

Over 200 WbLS samples with different materials and compositions were produced with a water fraction of 65\% or higher.
Their light yields were initially measured using a simple and conventional test system equipped with a photomultiplier tube (PMT) to efficiently evaluate many samples and understand the relationship between composition and light yield.
Based on the results, about 20 selected samples were encapsulated in small cells, and their light yields were subsequently measured using the fiber readout system.
The samples showed no significant loss in light yield after two months of exposure to air. 
The physical stability of these samples was also monitored over several months and no obvious changes were observed.
The stability will be continuously monitored to demonstrate year-scale performance required for neutrino beam operation.

\subsection{WbLS production}

We used pseudocumene (PC) as a solvent, 2,5-diphenyloxazol (PPO) as a fluorescent, 1,4-Bis(2-methylstyryl)benzene (Bis-MSB) as a wavelength shifter and Triton X-100 as a surfactant.
Since the Triton contains a benzene ring, it also works as a secondary solvent.
To enhance stability, sodium dodecyl sulfate (SDS) was added as a supplementary surfactant.
The procedure of producing WbLS is as follows.

\begin{enumerate}
   \item Measure the mass of the materials and mix them in a bottle.
   \item Stir the mixture using a hot stirrer.
   \item Break up undissolved clumps using an ultrasonic bath.
   \item Repeat steps (2) and (3) until the solution becomes transparent.
\end{enumerate}

\begin{figure}[htbp]
  \begin{center}
    \includegraphics[width=0.35\linewidth]{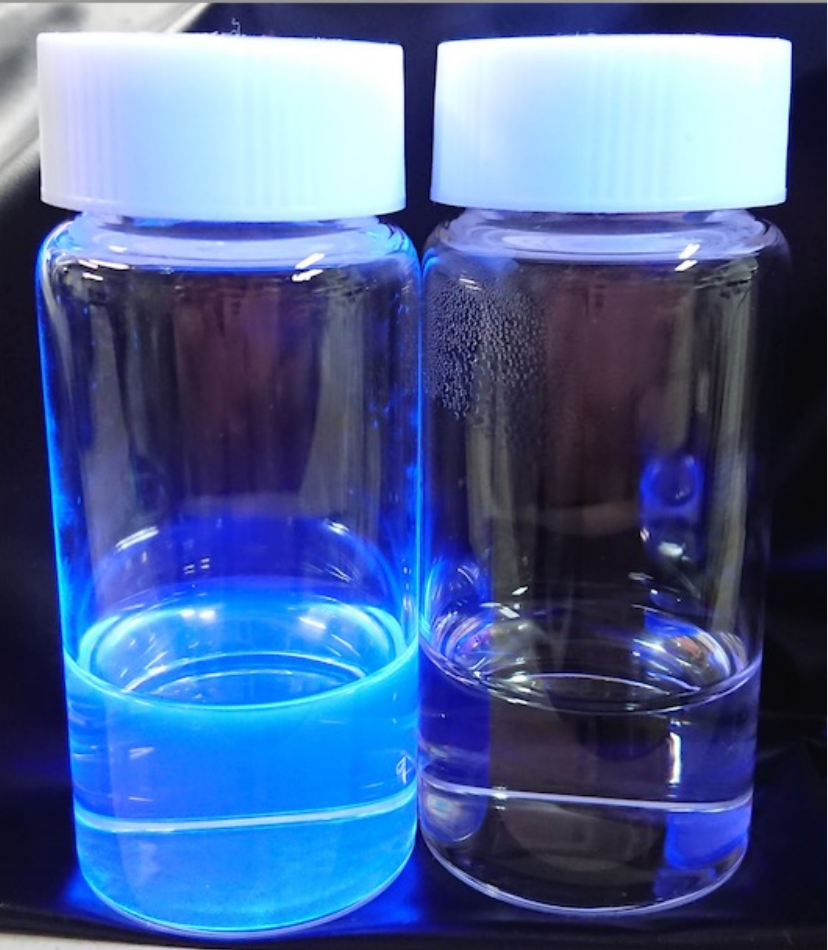}
    \caption{Examples of produced WbLS sample (left) and pure water (right) exposed to an ultraviolet light.}
    \label{wbls_photo}
  \end{center}
\end{figure}

We produced numerous WbLS samples with varying mixing ratios.
Although some samples failed due to powder sedimentation, many of them were improved by increasing the ratio of SDS or water.
During production, storage and measurement in this study, the WbLS samples were handled under ambient laboratory atmosphere; no nitrogen purging or inert-gas blanketing was applied at any stage.
Figure~\ref{wbls_photo} shows examples of the produced WbLS samples exposed to ultraviolet light.

\subsection{Light yield measurement with a PMT-based system}

\begin{figure}[htbp]
  \begin{center}
    \includegraphics[width=0.45\linewidth]{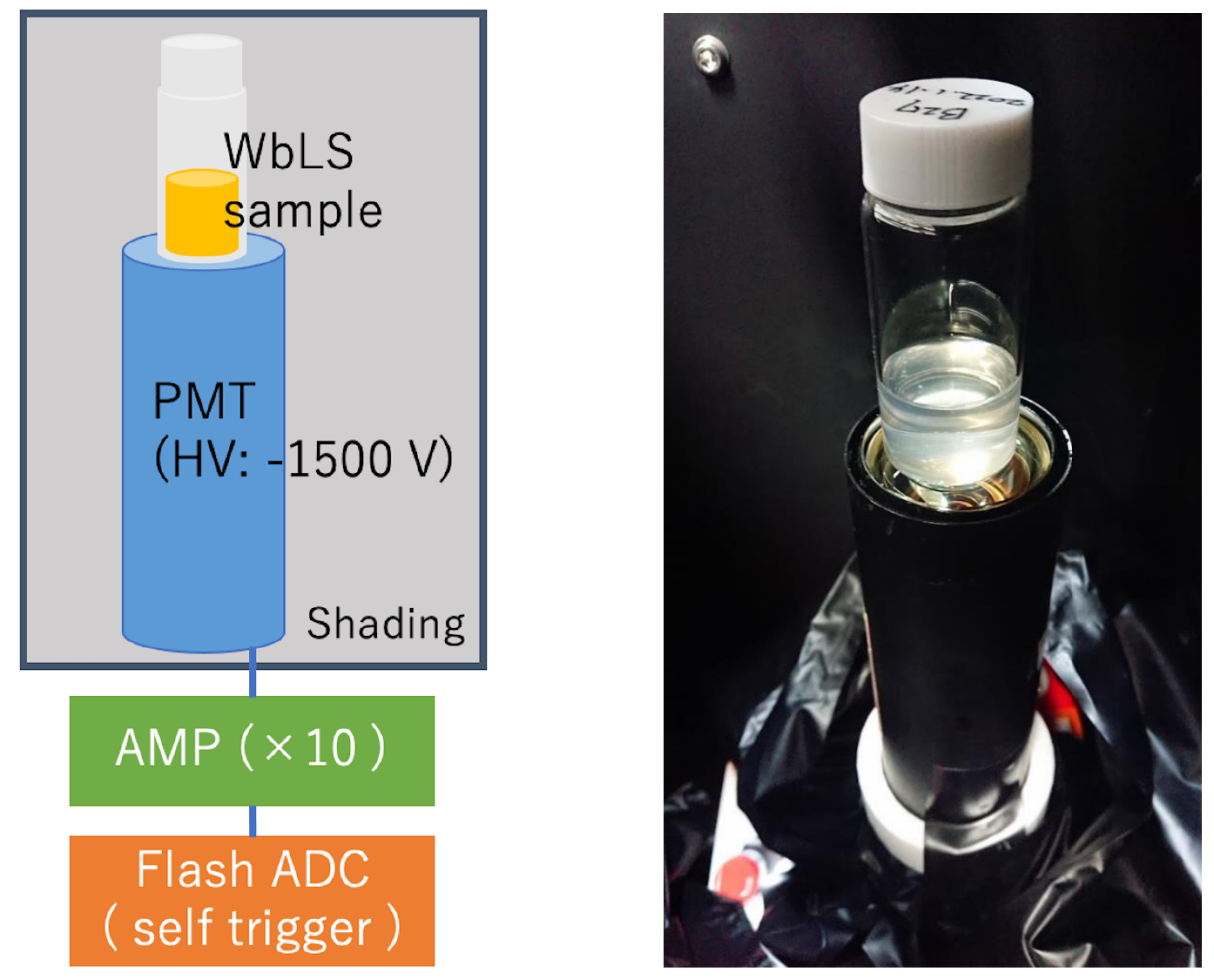}
    \caption{Setup of the light yield measurement with a PMT.}
    \label{test_pmt}
  \end{center}
\end{figure}

\begin{figure}[htbp]
    \begin{minipage}[b]{0.49\linewidth}
        \begin{center}
            \includegraphics[width=0.95\linewidth]{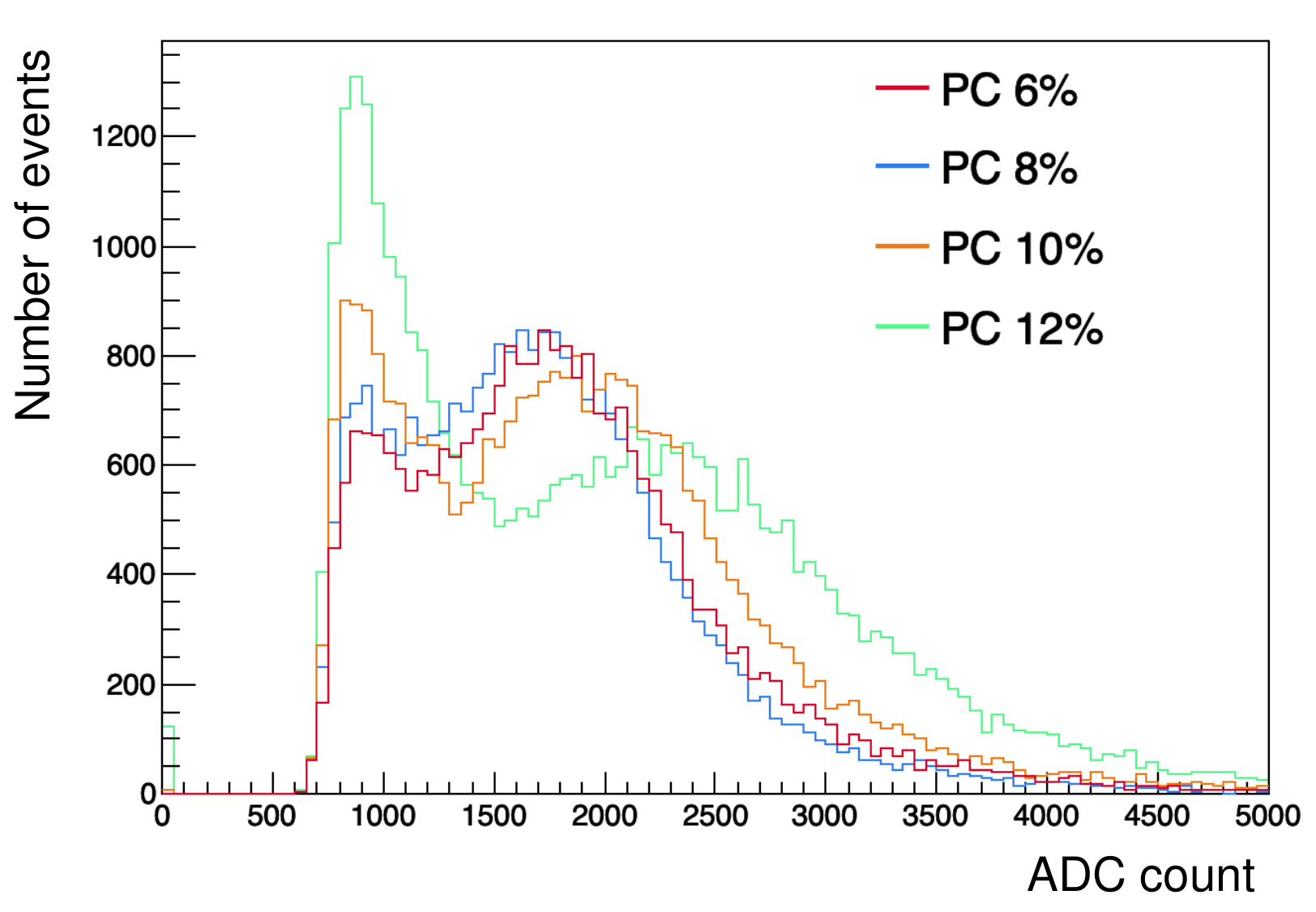}
            \subcaption{Various PC ratio}
        \end{center}
    \end{minipage}
    \begin{minipage}[b]{0.49\linewidth}
        \begin{center}
            \includegraphics[width=0.95\linewidth]{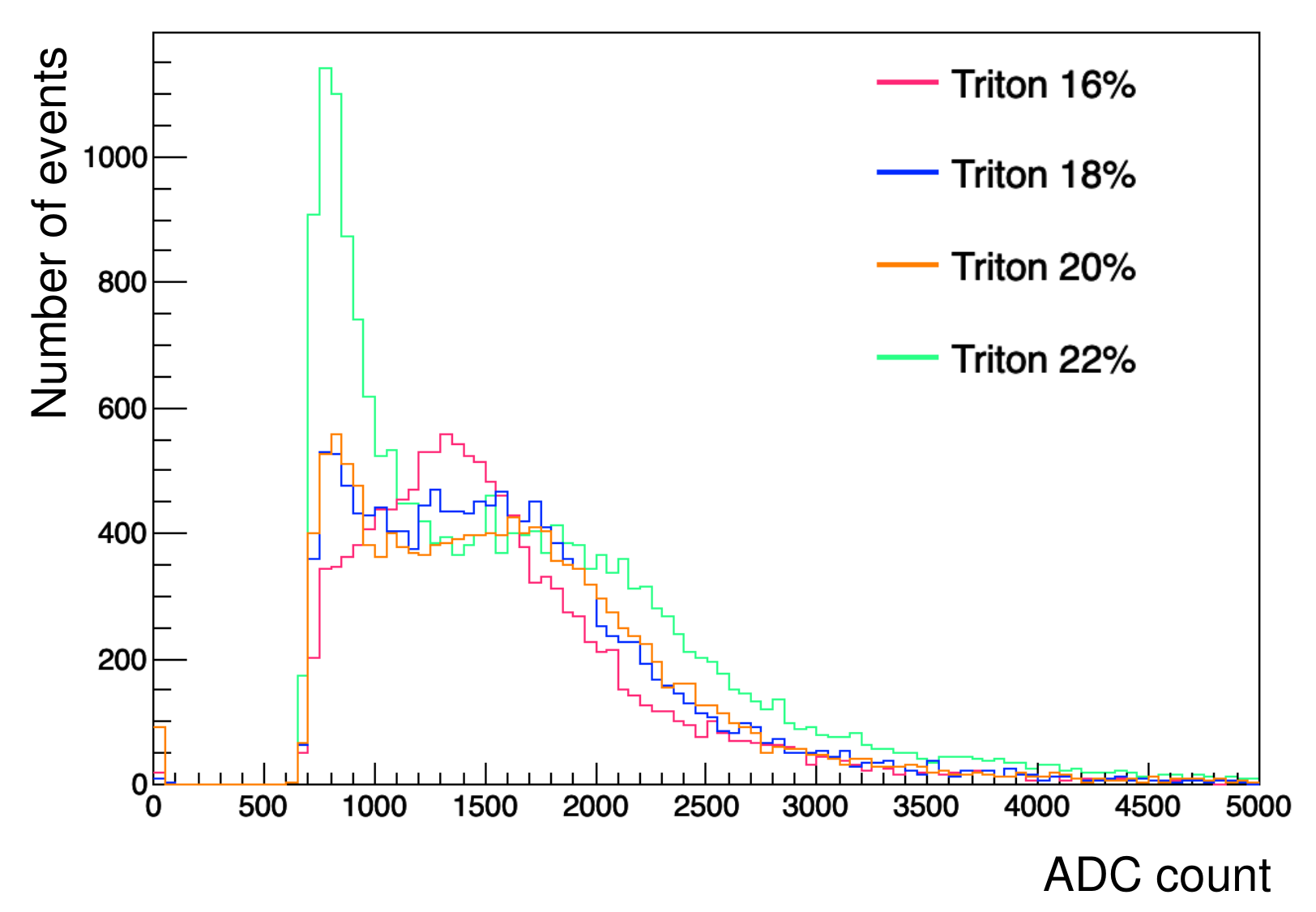}
            \subcaption{Various Triton ratio}
        \end{center}
    \end{minipage}
    \begin{minipage}[b]{0.49\linewidth}
        \begin{center}
            \includegraphics[width=0.95\linewidth]{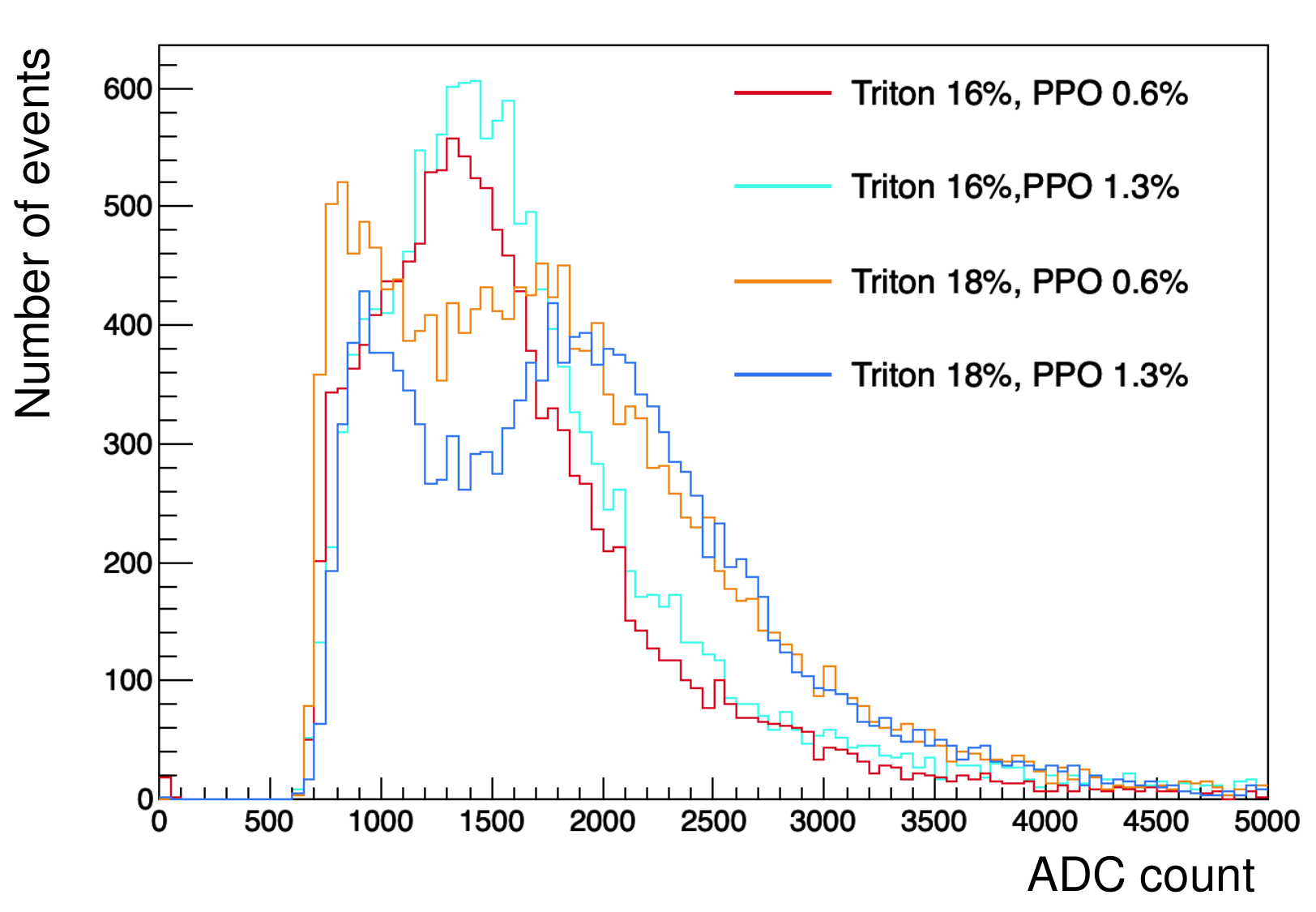}
            \subcaption{Various Triton and PPO ratio}
        \end{center}
    \end{minipage}
    \begin{minipage}[b]{0.49\linewidth}
        \begin{center}
            \includegraphics[width=0.95\linewidth]{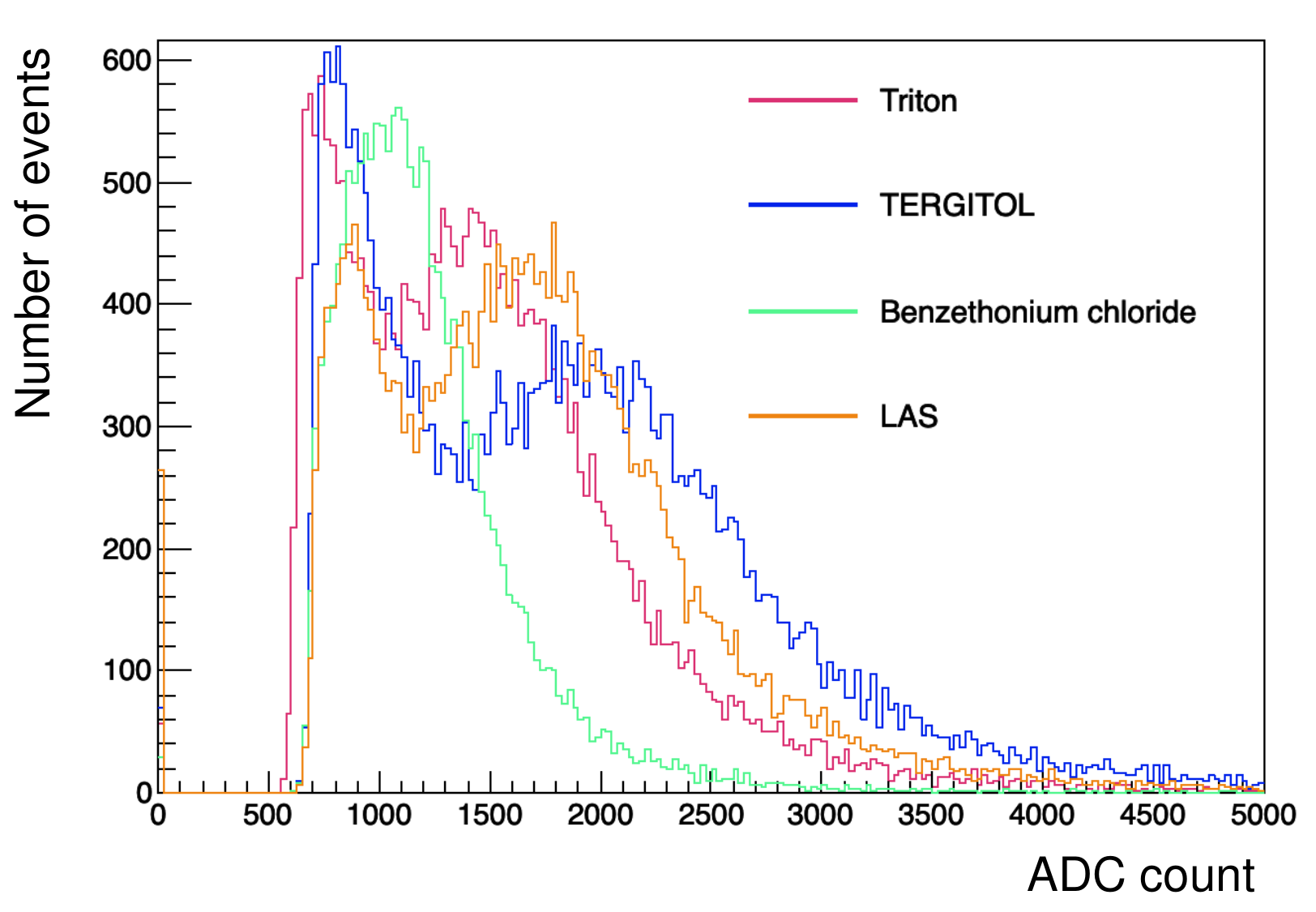}
            \subcaption{Various surfactants}
        \end{center}
    \end{minipage}    
    \caption{Result of the light yield measurements with a PMT-based system. The horizontal axis is the integral of the waveform.}
    \label{pmt_result}
\end{figure}

\begin{figure}[htbp]
  \begin{center}
    \includegraphics[width=0.85\linewidth]{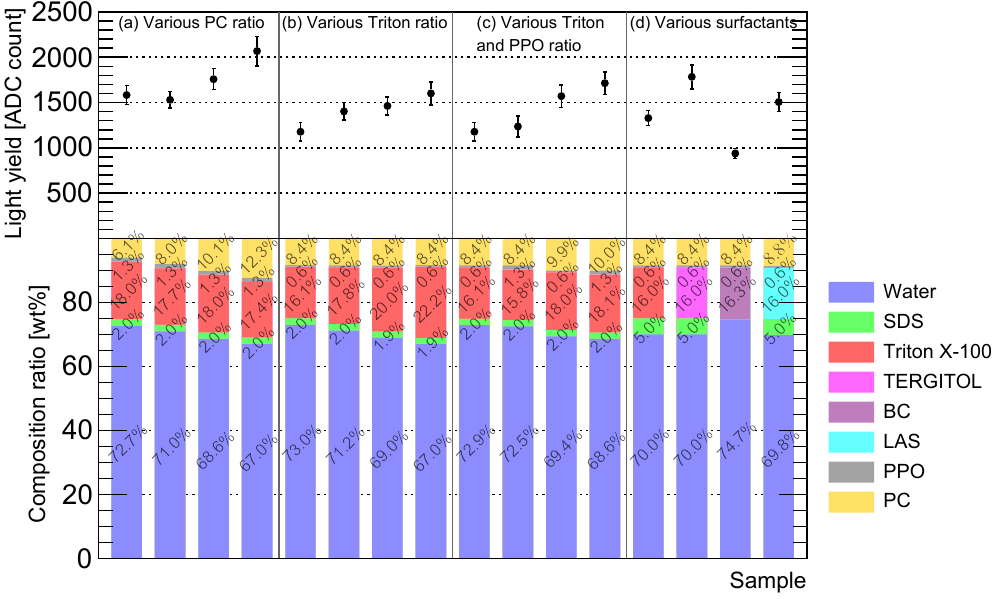}
    \caption{Composition and mean light yield of each WbLS sample measured with a PMT-based system. The error bars represent statistical errors.}
    \label{wbls_pmt_result}
  \end{center}
\end{figure}

The light yield of all the produced WbLS samples except for the failed ones was first measured using a PMT (Hamamatsu H6410).
The measurement setup is shown in Fig.~\ref{test_pmt}.
The bottles containing the WbLS samples were placed directly on the photocathode surface of the PMT.
The PMT was connected to a 12-bit, 62.5-MHz digitizer (CAEN DT5740), and cosmic-ray events were recorded with a self-trigger.
We investigated the relationship between the light yield and the mixing ratios of PC, Triton and PPO.
Figure~\ref{pmt_result}(a) shows the results for samples with 6\%, 8\%, 10\% and 12\% PC with other component ratios fixed at 2\% SDS, 18\% Triton, 1.3\% PPO, and 0.01\% Bis-MSB.
Figure~\ref{pmt_result}(b) presents the results for Triton concentrations of 16\%, 18\%, 20\%, and 22\%, keeping the other ratios constant at 2\% SDS, 8.4\% PC, 0.6\% PPO, and 0.01\% Bis-MSB.
Figure~\ref{pmt_result}(c) shows the effects of varying Triton (16\% and 18\%) and PPO (0.6\% and 1.3\%) concentrations, with fixed ratios of 2\% SDS and 0.01\% Bis-MSB.
The composition and mean light yield of each WbLS sample are summarized in Fig.~\ref{wbls_pmt_result}.
As expected, the mean light yield increased with higher concentrations of PC, Triton, or PPO. However, the light yield distribution became wider.
In addition, samples with higher Triton and PPO contents tended to be less stable in terms of their long-term chemical and optical stability.
In particular, the sample with 22\% Triton solidified into a gel, resulting in a broader distribution and a significant number of events with low light yield.

We also evaluated TERGITOL NP-10, benzethonium chloride (BC), and linear-alkyl-benzene-sulfonate (LAS) as alternative surfactants.
Since all of these compounds contain a benzene ring, they are expected to work as solvents as well as Triton. Among them, benzethonium chloride is a cationic surfactant, whereas the others including SDS are anionic surfactants. For this measurement, samples were made with surfactants containing benzene rings at a fixed concentration of 16\%. The sample containing TERGITOL exhibited the highest light yield as shown in Fig.~\ref{pmt_result}(d). However, the sample with TERGITOL tended to form a gel when the concentration exceeded 16\%, making it difficult to produce stable samples. Therefore, despite its high light yield, TERGITOL is not suitable for use in samples requiring higher concentrations.

\subsection{Light yield measurement with WLS fiber readout}
\begin{figure}[htbp]
  \begin{center}
    \includegraphics[width=0.75\linewidth]{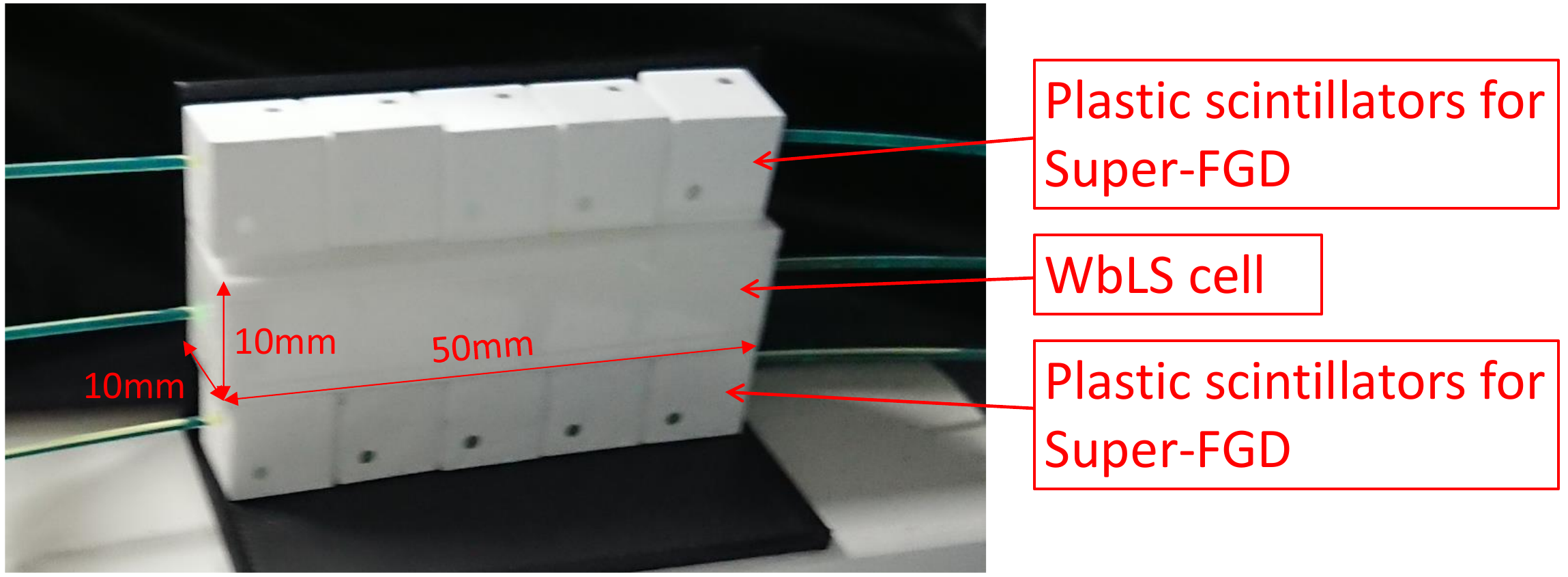}
    \caption{Setup of the light yield measurement with fiber readout.}
    \label{wbls_cell}
  \end{center}
\end{figure}

\begin{figure}[htbp]
    \begin{minipage}[b]{0.49\linewidth}
        \begin{center}
            \includegraphics[width=0.95\linewidth]{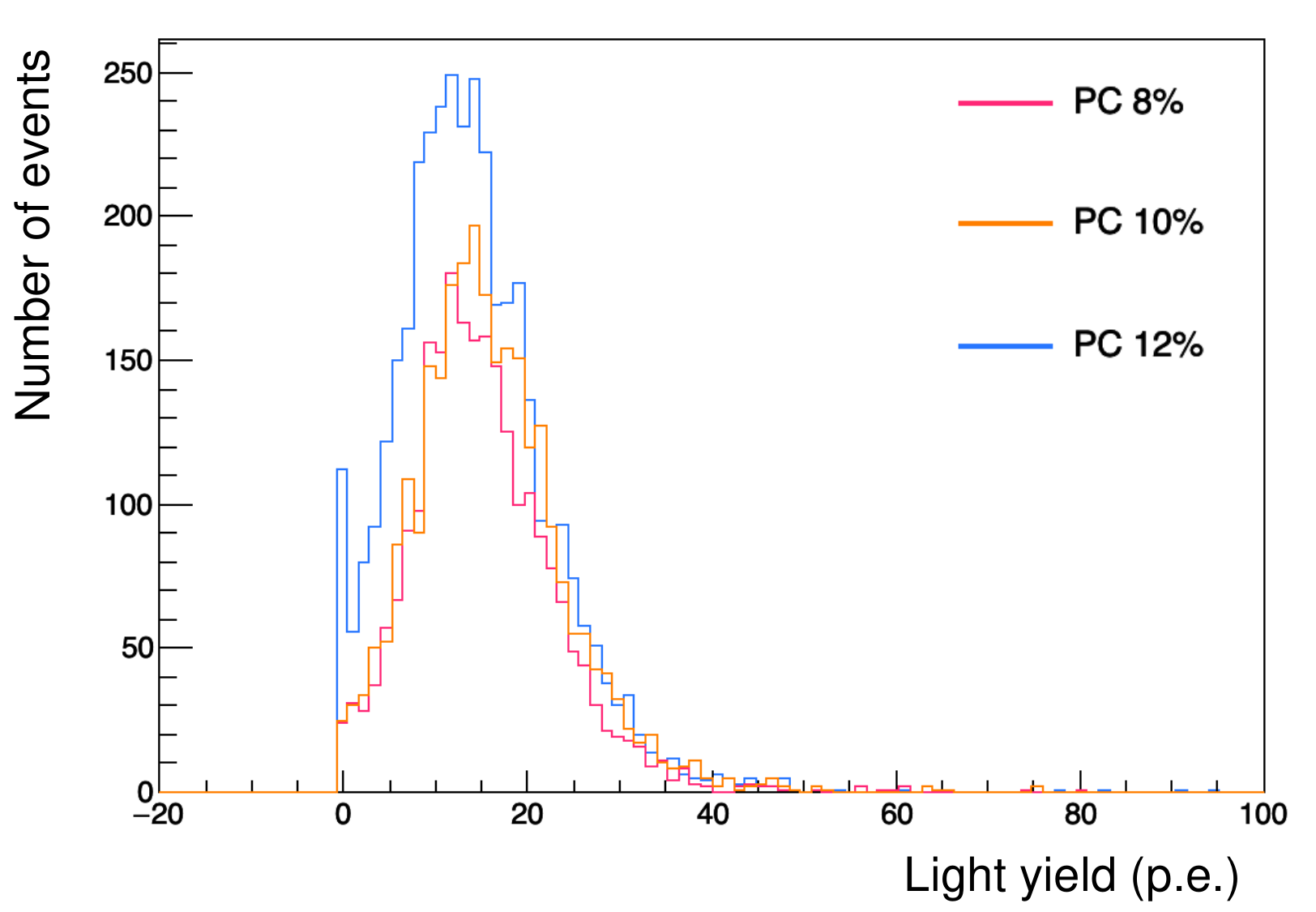}
            \subcaption{Various PC ratio}
        \end{center}
    \end{minipage}
    \begin{minipage}[b]{0.49\linewidth}
        \begin{center}
            \includegraphics[width=0.95\linewidth]{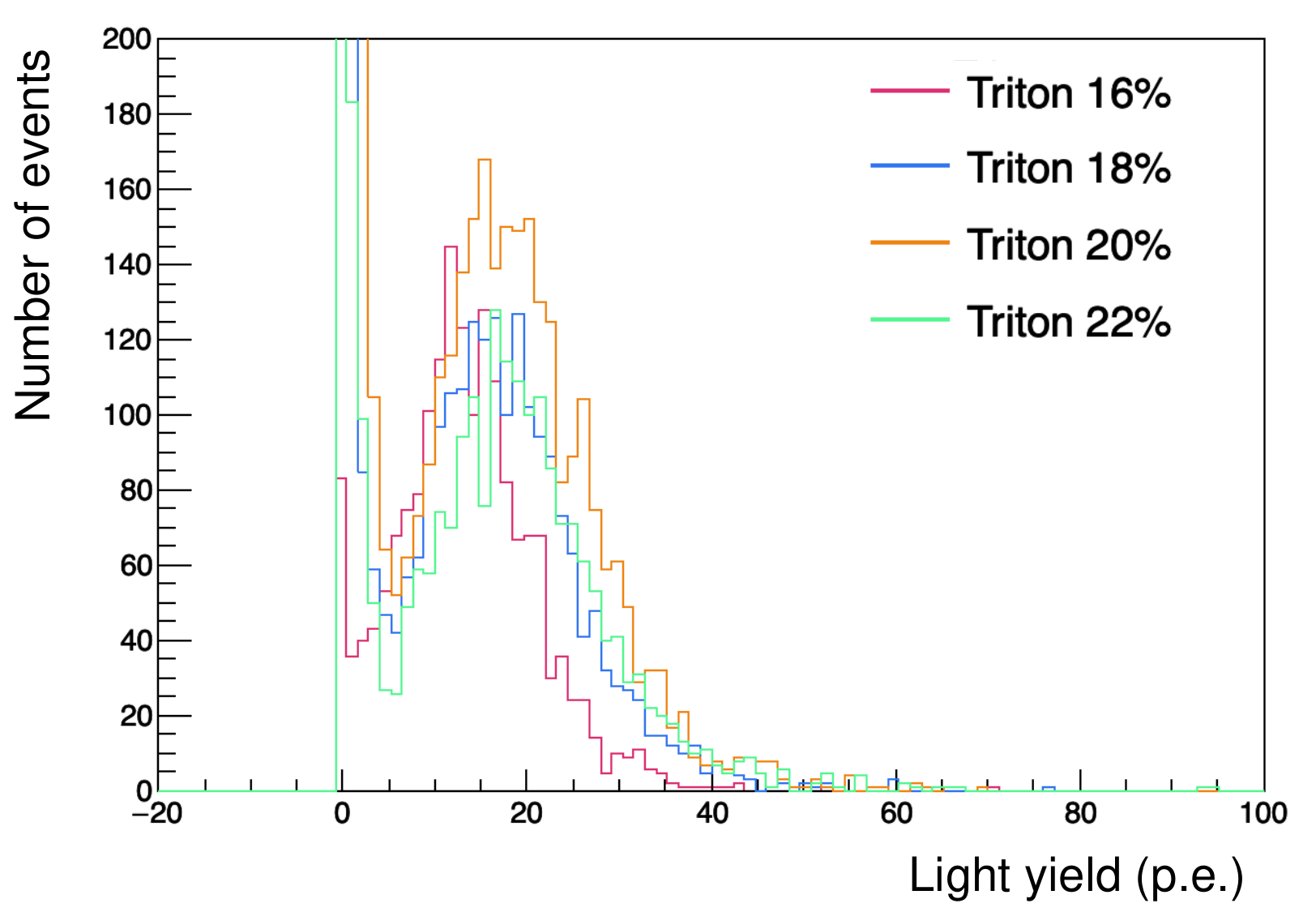}
            \subcaption{Various Triton ratio}
        \end{center}
    \end{minipage}
    \begin{minipage}[b]{0.49\linewidth}
        \begin{center}
            \includegraphics[width=0.95\linewidth]{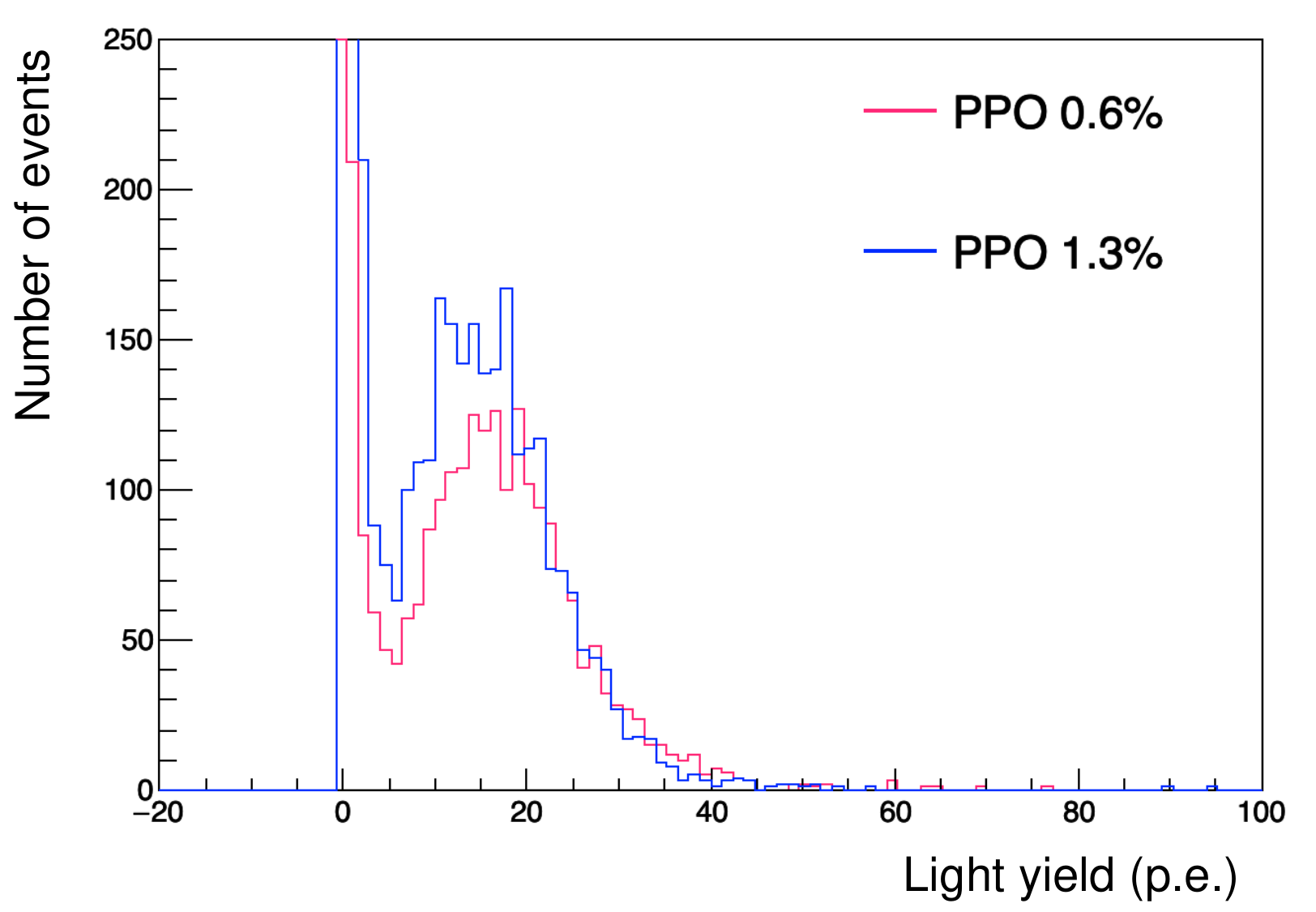}
            \subcaption{Various PPO ratio}
        \end{center}
    \end{minipage}
    \begin{minipage}[b]{0.49\linewidth}
        \begin{center}
            \includegraphics[width=0.95\linewidth]{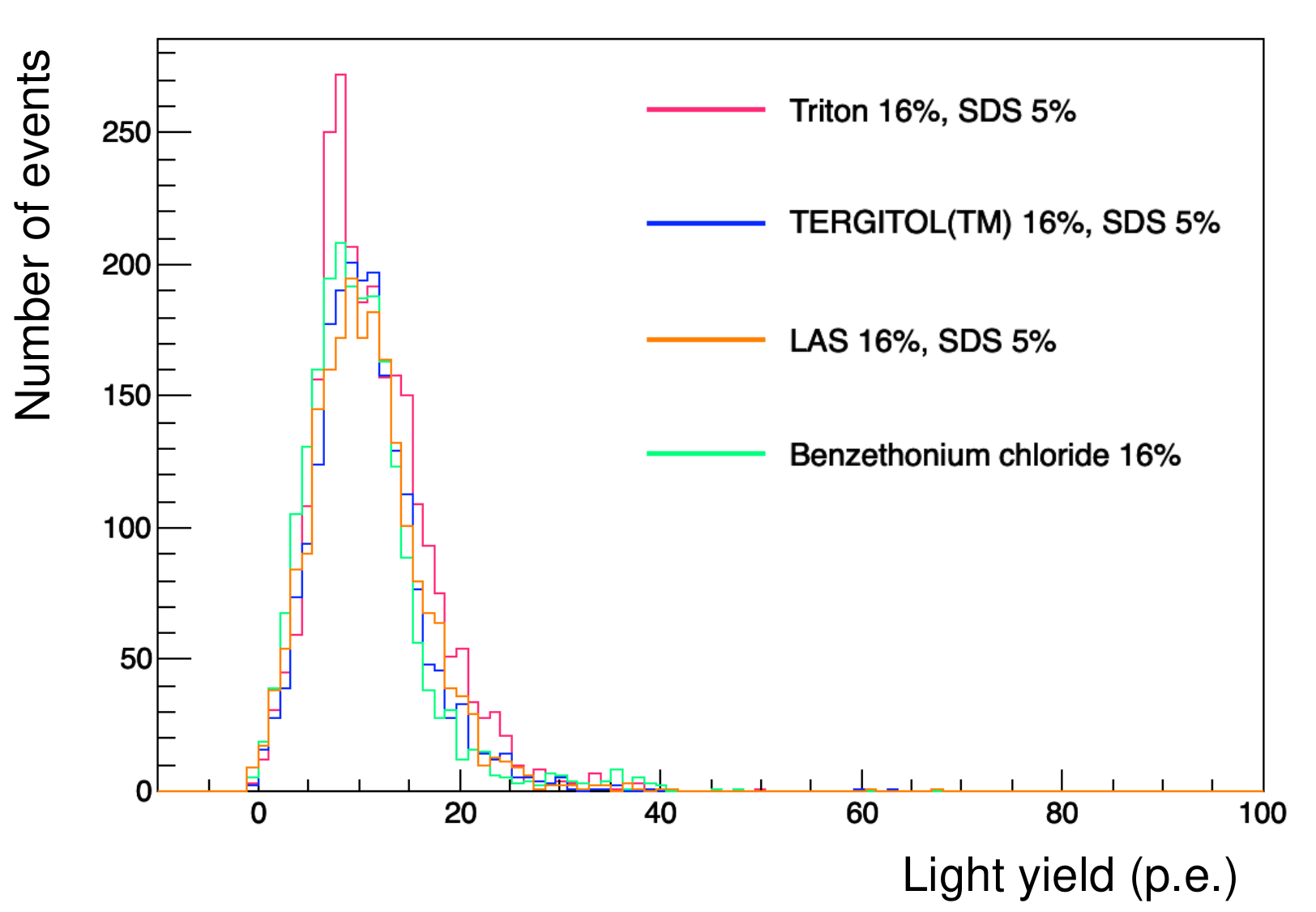}
            \subcaption{Various surfactants}
        \end{center}
    \end{minipage}    
    \caption{Result of the light yield measurements with fiber readout. The horizontal axis is the photon equivalent unit calibrated with dark noise signals.}
    \label{mppc_result}
\end{figure}

\begin{figure}[htbp]
  \begin{center}
    \includegraphics[width=0.85\linewidth]{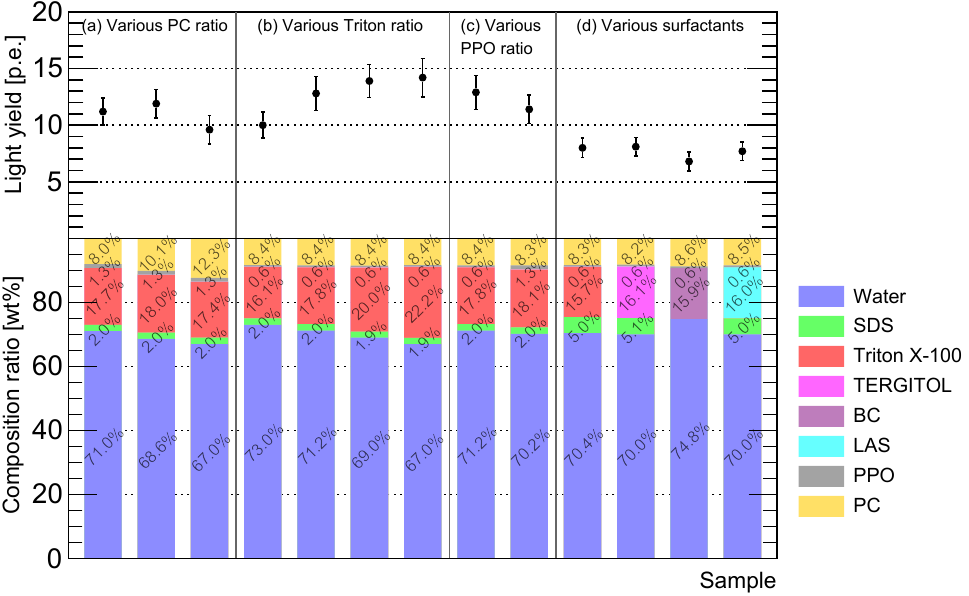}
    \caption{Composition and mean light yield of each WbLS sample measured with fiber readout. The error bars represent statistical errors.}
    \label{wbls_mppc_result}
  \end{center}
\end{figure}

The measurement with the PMT revealed the relationship between the light yield and the material composition of the WbLS.
However, since the PMT and fiber readout systems have different spectral sensitivities, and the optical path lengths differ between the full bottle and the small cell, a direct comparison is not straightforward.
To obtain results more relevant to the actual detector configuration, selected WbLS samples were further tested using the fiber readout in small cells.
The measurement setup is shown in Fig.~\ref{wbls_cell}.
The WbLS samples were encapsulated in 1~cm $\times$ 1~cm$ \times$ 5~cm white acrylic cells with 1~mm wall thickness, fabricated using a 3D printer.
The cell had a hole through which a WLS fiber (Kuraray Y11(200)M 1.0mmD BSJ) was inserted into the WbLS volume.
After insertion, the hole was sealed with optical cement (Eljen EJ-500) to prevent a leakage of the WbLS.
The fiber end was coupled to an SiPM (Hamamatsu  S10362-13-050C), and the SiPM signal waveform was recorded by a digitizer (CAEN DT5740).
Plastic scintillators identical to those used in the T2K SuperFGD were placed at the top and bottom of the WbLS cell.
Cosmic-ray events were selected by requiring coincident signals in both top and bottom scintillators.
The measured ADC values were linearly converted to the number of photoelectrons using the pedestal and the one-photoelectron equivalent ADC value obtained from the SiPM dark noise signals, and this converted value was used as the observable of light yield.
Figure~\ref{mppc_result}(a)(b)(c) show the results for different ratios of PC, Triton, and PPO.
The light yield increased with the Triton ratio, but showed little dependence on the PC ratio.
Interestingly, increasing the PPO concentration from 0.6\% to 1.3\% resulted in a decrease in light yield.
As shown in Fig.~\ref{mppc_result}(d), changing the surfactant material did not significantly affect the light yield.
The composition and mean light yield of each WbLS sample are summarized in Fig.~\ref{wbls_mppc_result}.
These results indicate that the light yield trends observed with the fiber readout differ substantially from those obtained with the measurement using a PMT.
The typical mean light yield observed with the fiber readout was 10–14 p.e.
However, this measurement was performed using only a single fiber readout and employed cells with dimensions different from those of the actual detector.
In addition, cosmic rays entered the setup from various angles, and optical cross-talk from the adjacent plastic scintillators may have affected the results.
Therefore, the light yield in the actual detector configuration was evaluated in the beam test described in Sec.~\ref{sec_beamtest}.
Nevertheless, this measurement marked the first successful detection of WbLS scintillation light using the WLS fiber readout system.

\section{Positron beam test}\label{sec_beamtest}
The measurement using cosmic rays demonstrated the feasibility of the WLS fiber readout of the WbLS light signal.
In addition, several promising WbLS samples were identified through the measurement.
Using these selected WbLS samples, we fabricated a prototype of the WbLS tracking detector and evaluated its performance with a 500~MeV positron beam at the Research Center for Accelerator and Radioisotope Science (RARiS), Tohoku University.
The goals of the beam test are to demonstrate the tracking capability and to evaluate the light yield variations between cells as well as position dependence within a cell.

\subsection{Setup}

\begin{figure}[htbp]
  \begin{center}
    \includegraphics[width=0.65\linewidth]{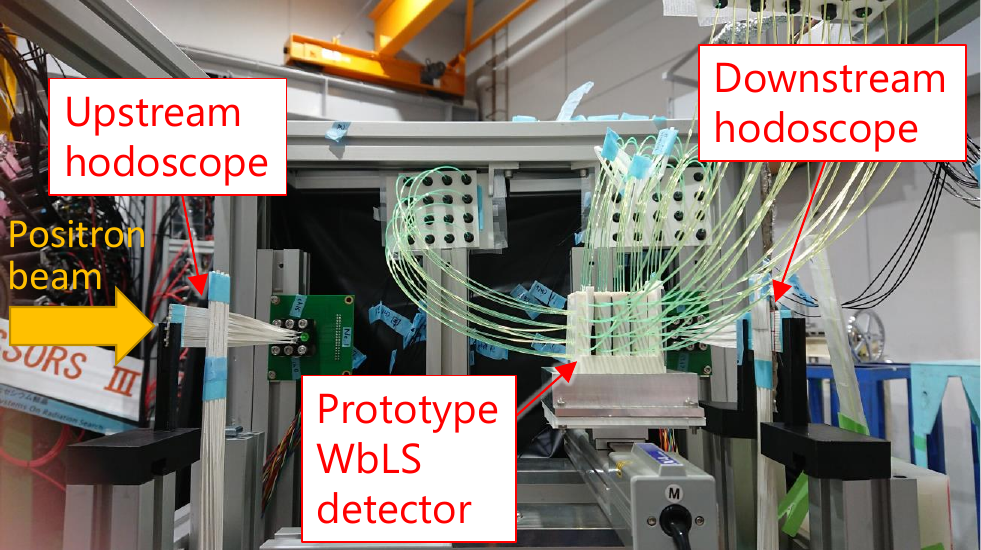}
    \caption{Setup of the positron beam test at RARiS.}
    \label{elph_setup}
  \end{center}
\end{figure}

The overall experimental setup of the positron beam test is shown in Fig.~\ref{elph_setup}.
The prototype WbLS detector was placed on the beamline of the GeV-$\gamma$ room at RARiS.
Two pairs of horizontal and vertical hodoscopes were placed upstream and downstream of the WbLS detector to provide triggers and precise positional information of the positrons.
The coordinate system is defined with the Z-axis aligned with the beam direction, and the X- and Y-axes defined as the horizontal and vertical directions perpendicular to the beam, respectively.

\subsubsection{Beamline at RARiS}

An electron beam is first accelerated up to 200~MeV at a maximum by a linear accelerator.
Then, the electron beam is injected into a synchrotron called Booster-STorage (BST), and accelerated up to 1.3~GeV at a maximum.
A carbon wire in orbit of the accelerated electrons produces high-energy $\gamma$-rays via bremsstrahlung.
These $\gamma$-rays subsequently produce electron–positron pairs in a tungsten target.
The momentum of the resulting electrons or positrons is analyzed using a bending magnet.
A 500~MeV/c positron beam was selected for this test.
The beam width is approximately 7~mm in both X and Y directions and was monitored throughout the test. The beam rate is approximately 3~kHz with a maximum duty cycle of 0.63.

\subsubsection{Prototype WbLS tracking detector}

\begin{figure}[htbp]
    \begin{center}
        \includegraphics[width=0.8\linewidth]{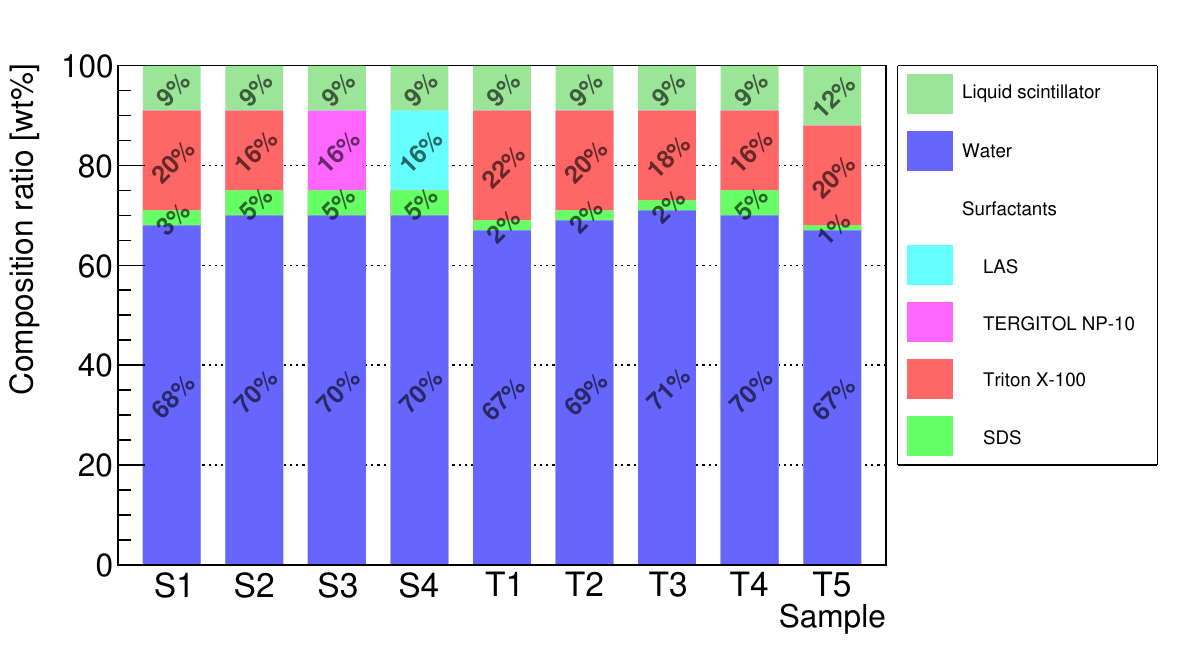}
        \caption{
            Compositions of the WbLS used in prototype detectors for the beam test. 
        }
        \label{fig:WbLS_beamtest}
    \end{center}
\end{figure}

The prototype WbLS tracking detectors consisted of five layers.
Each layer had a $4\times4$ array of cells with internal dimensions of 1~cm $\times$ 1~cm$ \times$ 1~cm.
A specially designed white acrylic vessel with 1~mm wall thickness optically separated the cells and had holes for the fiber readout.
This vessel was fabricated using a high-precision 3D printer.
The upstream layer had a three-directional fiber readout structure, with 4$\times$4 fibers (Kuraray Y11(200)M 1.0mmD BSJ) at the front and four fibers each at the top and side.
Each of the remaining four layers was equipped with two-directional readouts, with four fibers each at the top and side.
The five layers were arranged along the beam direction.
Light-shielding sheets were sandwiched between the layers to prevent optical cross-talk between the layers. Each layer was filled with a WbLS sample of different compositions, as shown in Fig.~\ref{fig:WbLS_beamtest}.
Two prototype detectors were made.
One used WbLS samples S1–S4, which varied in the type of surfactant used.
The other used samples T1–T5, all of which used Triton X-100 as the surfactant but differ in material ratios.
In both cases, SDS was added to improve the stability of the WbLS.
The S1 and T1 samples were filled in the upstream layer with the three-directional fiber readout, and other samples were filled in the latter layers which had only two-directional fibers.
Each fiber was coupled to an SiPM (Hamamatsu S10362-13-050C), with a total of 56 readout channels.
Signals from the SiPMs were digitized by a NIM EASIROC module\cite{bib:easiroc}.

\subsubsection{Hodoscopes}

Each hodoscope consists of 16 scintillating fibers (Kuraray SCSF-78SJ) arranged in a linear array.
The fiber has a square cross-section of 1.5$\times$1.5~mm$^2$, and are manually coated with a TiO$_2$-based reflective paint (Eljen EJ-510) approximately 0.1~mm thick. A pair of hodoscopes is perpendicularly stacked for tracking. The scintillation light is detected by a 4 $\times$ 4 SiPM array (Hamamatsu S13361-3050AE-04), coupled to the fiber bundle.
The SiPMs were read out using another EASIROC module, which was synchronized with the module used for the WbLS prototype detector.
Data from all channels were recorded when two or more channels in the upstream hodoscope had hits.

\subsection{Analysis}
\begin{figure}[htbp]
    \begin{minipage}[b]{0.325\linewidth}
        \begin{center}
            \includegraphics[width=0.97\linewidth]{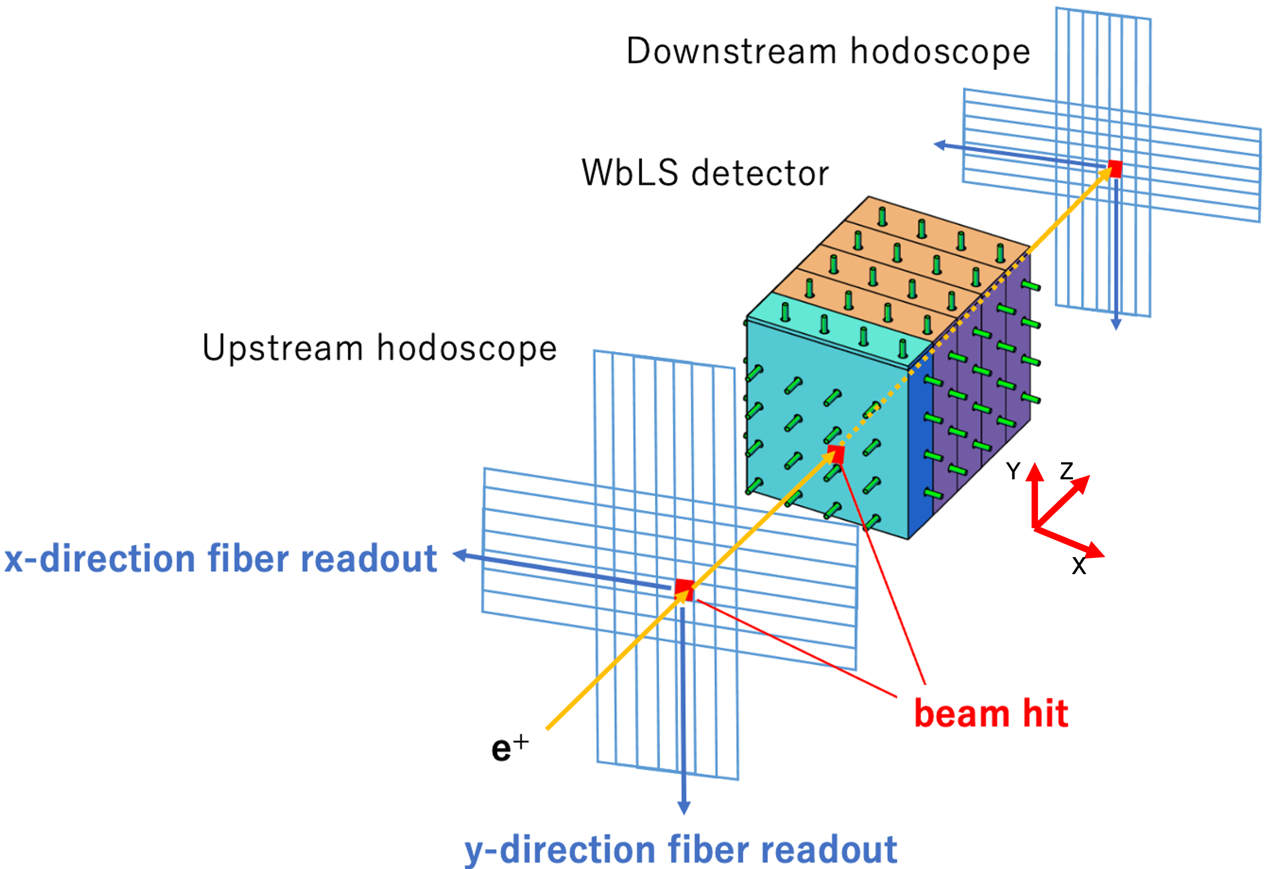}
            \subcaption{}
            \label{hodoscope_hit_a}
        \end{center}
    \end{minipage}
    \begin{minipage}[b]{0.325\linewidth}
        \begin{center}
            \includegraphics[width=0.97\linewidth]{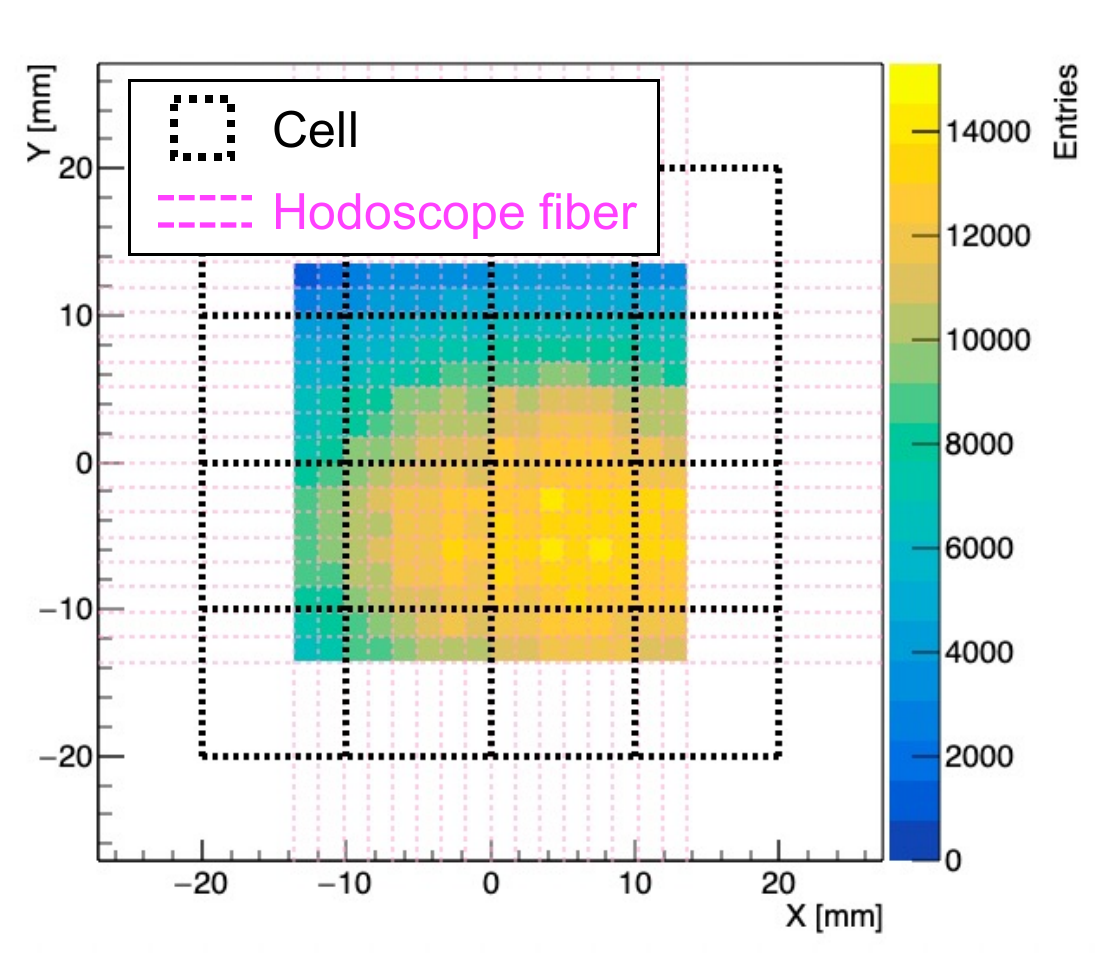}
            \subcaption{}
            \label{hodoscope_hit_b}
        \end{center}
    \end{minipage}
    \begin{minipage}[b]{0.325\linewidth}
        \begin{center}
            \includegraphics[width=0.97\linewidth]{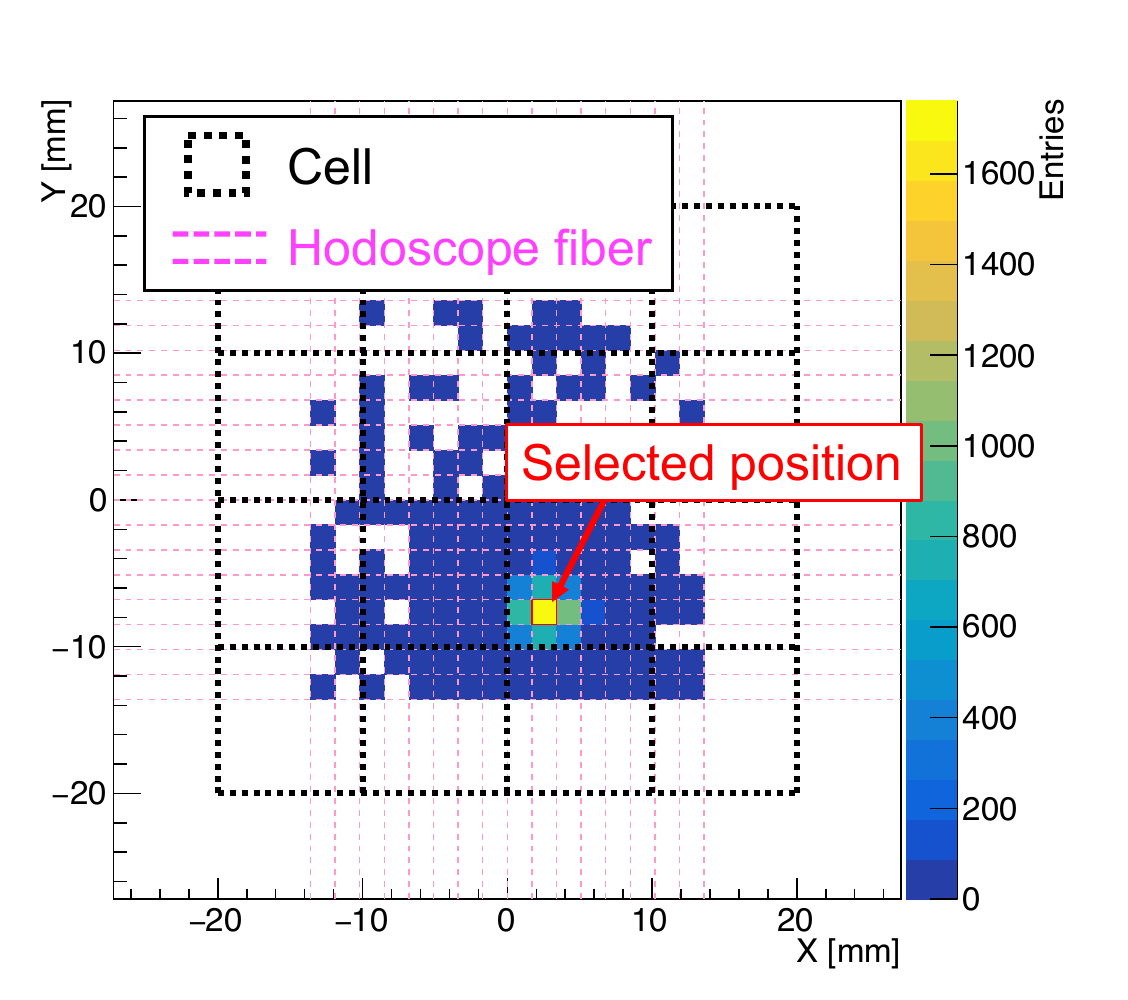}
            \subcaption{}
            \label{hodoscope_hit_c}
        \end{center}
    \end{minipage}
    \caption{Event selection using beam position measured by hodoscopes. (a) Conceptual drawing of the beam position measurement by hodoscopes. (b) Hit positions at the upstream hodoscopes.
    (c) Hit map at downstream hodoscopes with upstream hit selection.
    }
    \label{hodoscope_hit}
\end{figure}

The positron trajectory through the WbLS detector was reconstructed based on the hit positions in the hodoscopes.
As illustrated in Fig.~\ref{hodoscope_hit_a}, we selected events where only one fiber in both the horizontal and vertical hodoscope layers recorded a hit.
Figure~\ref{hodoscope_hit_b} shows the hit distribution for the upstream hodoscope, and Fig.~\ref{hodoscope_hit_c} shows corresponding downstream hits for a selected upstream position.
Many events had the same hit positions in the upstream and downstream hodoscopes.
We selected the straight-through track event by choosing these events.
Events with different hit positions were expected to come from the scattering in the prototype WbLS detector and were not used for the further analysis.

\subsection{Results}

\subsubsection{Positron track detection} 
\begin{figure}[htbp]
    \begin{center}
        \includegraphics[width=0.65\linewidth]{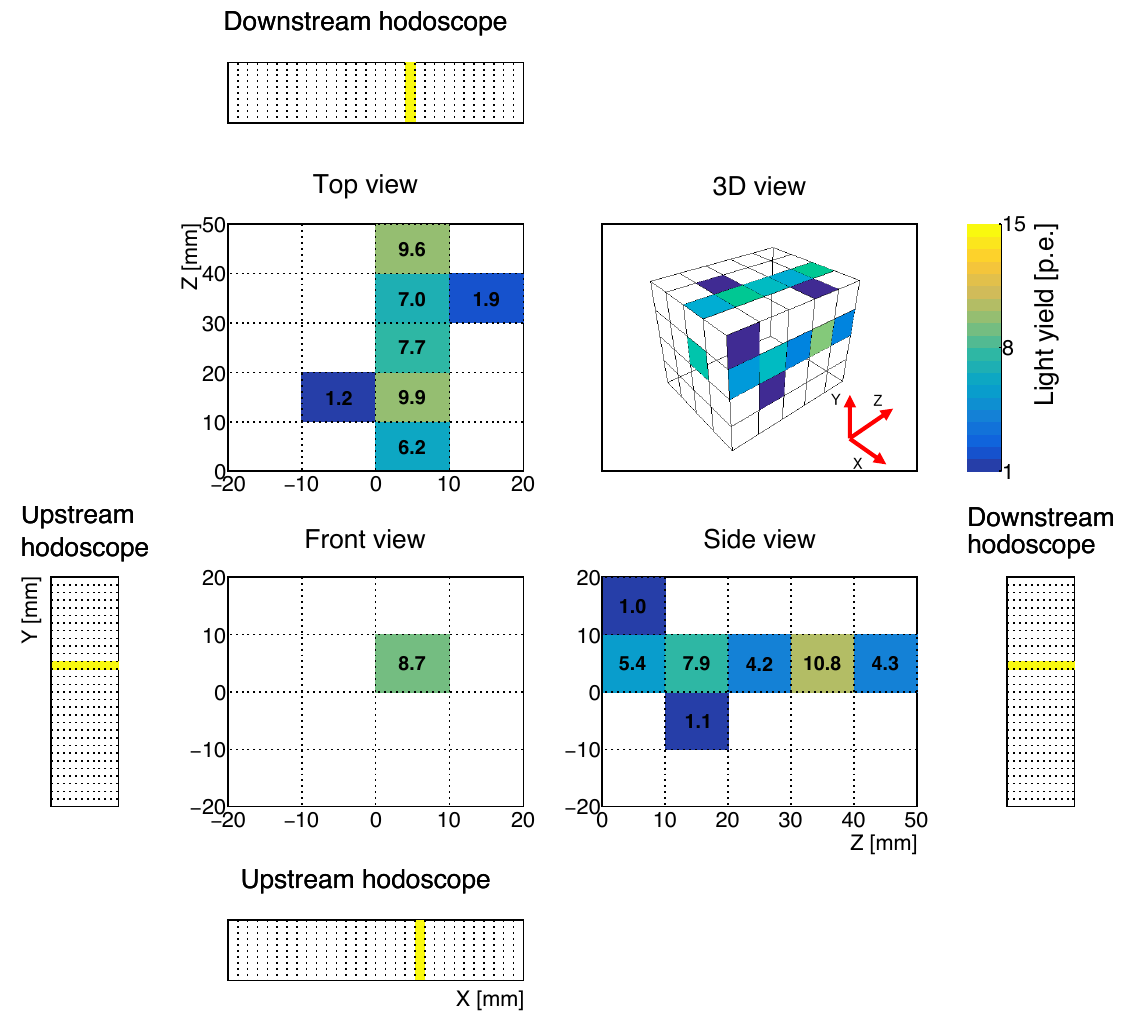}
        \caption{
        Plane views and 3D view of an event from the positron beam test data. 
        Each view shows the detected light yield from one of the three readout planes of the prototype WbLS detector. 
        The positron beam passed along the Z direction. 
        }
        \label{eventdisplay}
    \end{center}
\end{figure}

We succeeded in detecting
positron tracks in the prototype WbLS detector.
Figure~\ref{eventdisplay} shows an event observed in the positron beam test.
The track in the prototype WbLS detector was observed in the same position as hits in the upstream and downstream hodoscopes.

\subsubsection{Light yield of prototype WbLS detector} \label{sec_beamtest_ly}
\begin{figure}[htbp]
\begin{center}
    \begin{minipage}{0.4\linewidth}
        \centering
        \includegraphics[width=0.9\linewidth]{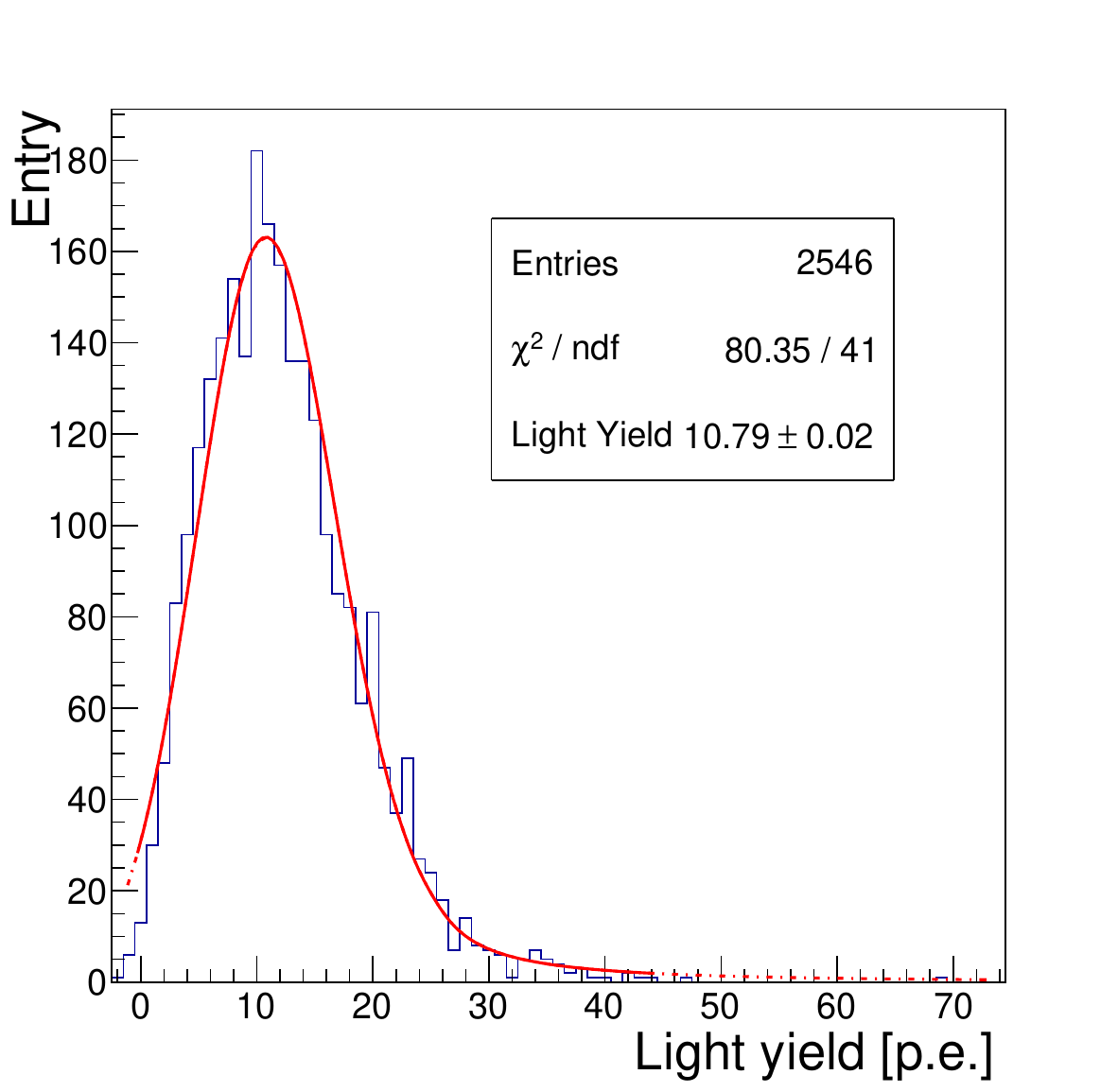}
        \subcaption{Sample T1}
    \end{minipage}
    \begin{minipage}{0.4\linewidth}
        \centering
        \includegraphics[width=0.9\linewidth]{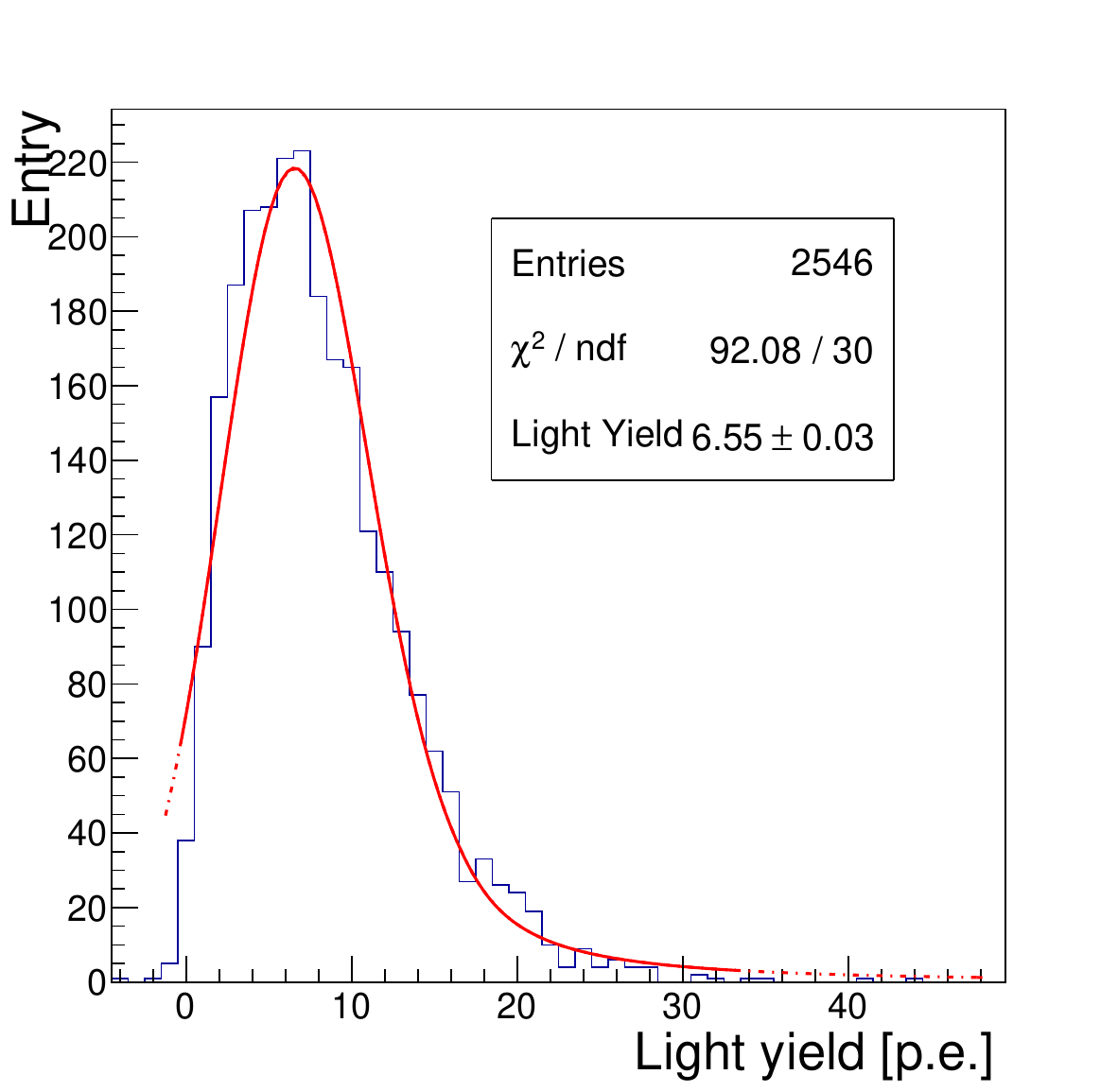}
        \subcaption{Sample T2}
    \end{minipage}
    \caption{Light yield distribution of the prototype WbLS detector. 
    The peak is fitted with a Landau distribution convoluted with a Gaussian distribution and most probable value is used as light yield.
    (a) A distribution of Sample T1 which has three-directional fiber readout. 
    (b) A distribution of Sample T2 which has only two-directional fiber readout.
    }
    \label{fig:lightyield_ex}
\end{center}
\end{figure}

The light yield is a fundamental characteristic of the detector and is important for detecting the particle track and identifying particles such as protons and pions from neutrino interactions.
Events in which positrons passed through the center of cells were selected for light yield evaluation.
Figure~\ref{fig:lightyield_ex} shows the distribution of the sum of the light yield in three-directional (two-directional) fiber readouts for samples T1 (T2) after the event selection.
The distribution was fitted with a Landau distribution convoluted with a Gaussian distribution, and the most probable value of the fitted function was used as the light yield. 
In this analysis, the $2\times2$ cells in the centers of each layer were used.

\begin{figure}[htbp]
    \begin{center}
         \includegraphics[width=0.7\linewidth]{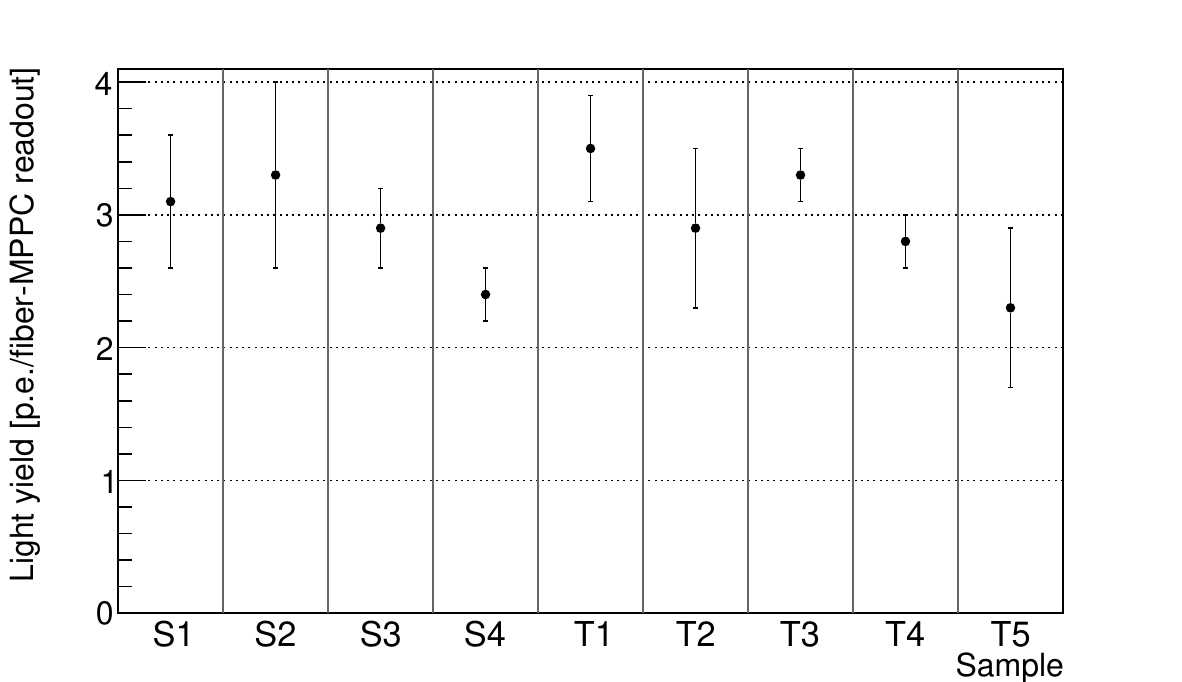}
         \caption{
            The light yield of prototype WbLS detector. The error bars represent the quadrature sum of the statistical uncertainty and the standard deviation among the four cells.
         }
         \label{fig:lightyield}
    \end{center}
\end{figure}

The mean light yields for four cells per fiber readout and errors are shown in Fig.~\ref{fig:lightyield}.
These errors include the statistical uncertainty and the standard deviation for four cells.
There was about 20\% variation of light yield among four cells at maximum
which results in much larger errors than the statistical uncertainties.
The typical light yield by positron passage was less than 4~p.e. per fiber readout.
The mean energy deposit of a 500~MeV positron per cell was estimated to be 1.45~MeV using Geant4 simulation\cite{geant4_1, geant4_2, geant4_3}.
Therefore, the maximum light yield corresponded to 2.4~p.e./MeV per fiber readout. To achieve more than 99\% efficiency in MIP detection at each fiber, an approximately 3.4 times larger light yield would be required.
There were no significant differences in the light yield per fiber readout between samples with three-directional fibers and those with two-directional fibers, implying a low scintillation photon collection efficiency in a cell.

\subsubsection{Non-uniformity of light yield in cell} 
The light yield at each fiber can vary with the position of scintillation photon generation due to different distances to the readout fibers. 
Non-uniformity was calculated as the standard deviation divided by the mean of the light yield at each position.
The $4\times4$ hit position in the center of the cell was used to avoid the events in which positrons pass through the cell wall. 

\begin{figure}[htbp]
\begin{center}
    \includegraphics[width=0.7\linewidth]{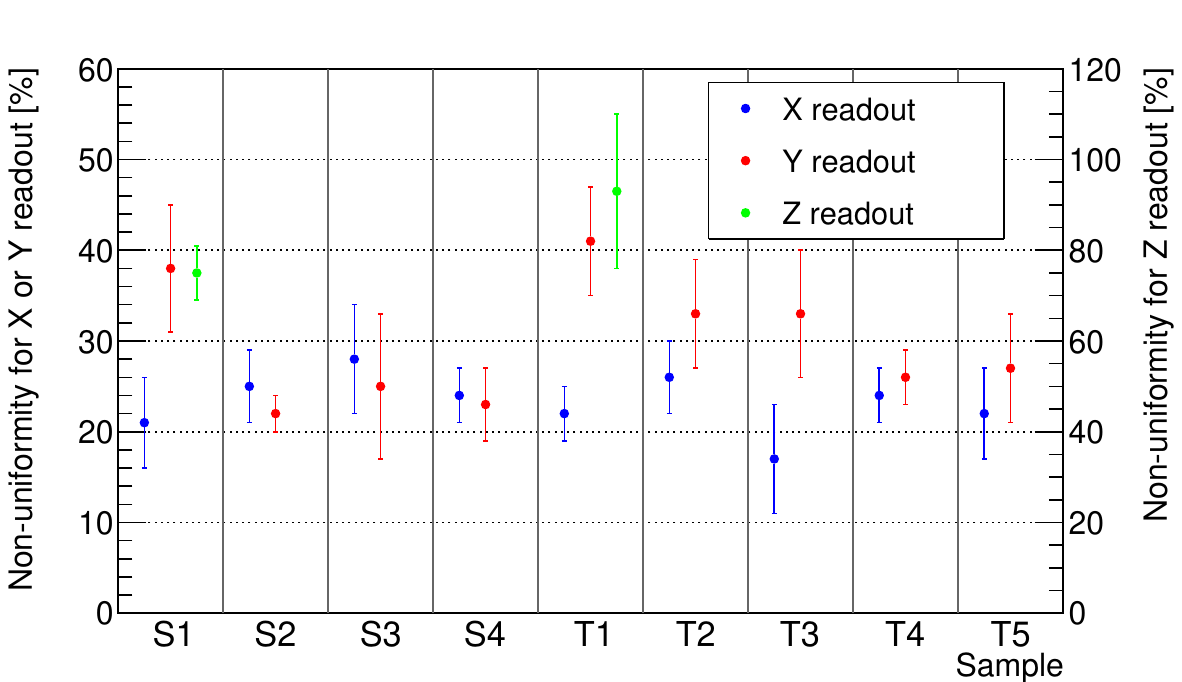}
    \caption{Non-uniformity of the light yield in the cell for each sample. The vertical axis scale for Z readout is twice compared to that of X or Y readout. The Z readout values are available only for samples S1 and T1, which were placed in the first layer of the prototype detector that had three-directional fiber readout.
    }
    \label{fig:non-uniformity}
\end{center}
\end{figure}

\begin{figure}[htbp]
\begin{center}
    \begin{minipage}{0.45\linewidth}
        \includegraphics[width=0.9\linewidth]{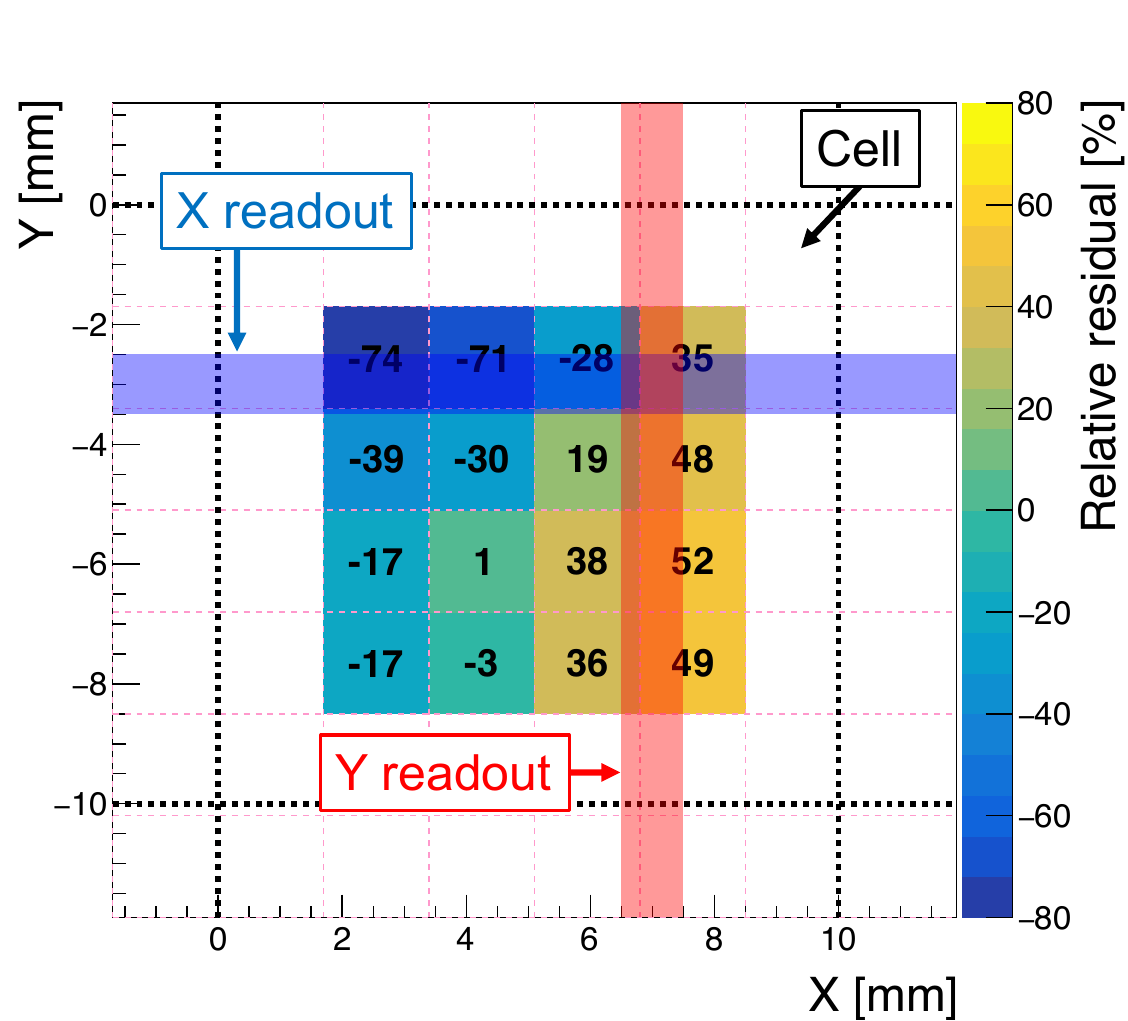}
        \subcaption{Sample T2}
        \label{fig:position_dependence_a}
    \end{minipage}
    \begin{minipage}{0.45\linewidth}
        \includegraphics[width=0.9\linewidth]{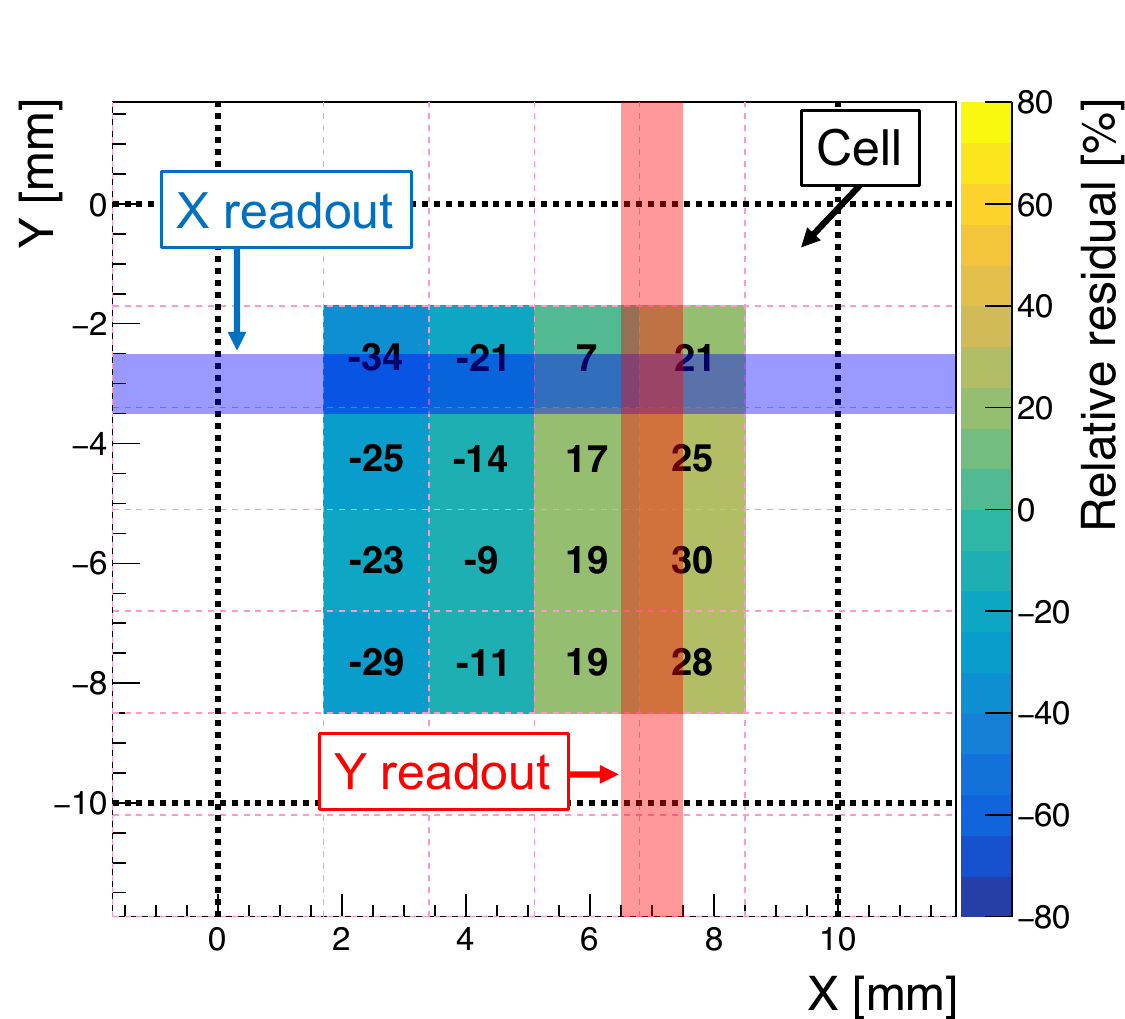}
        \subcaption{Sample T4}
        \label{fig:position_dependence_b}
    \end{minipage}
    \caption{Position dependence of the light yield from the Y readout. Relative residual from the mean light yield at each hit position at hodoscopes is shown. 
    }
    \label{fig:position_dependence}
\end{center}
\end{figure}

Figure~\ref{fig:non-uniformity} shows the non-uniformity of the light yield in the X, Y, and Z directions.  There were typically 20 - 30\% variations in the light yield in the X and Y readout. By contrast, the variations in the Z readout were about twice as large. Note that Z readout fibers were inserted only in the upstream layer, and not in other layers.  
Therefore, the Z readout data were only available for S1 and T1 samples. 
In the Z direction, positrons passed parallel to the readout fibers, so the distance between the positron track and the fiber directly determined the distance between the generated photon to the fiber. In contrast, in the X and Y directions, positrons crossed the fibers perpendicularly, and even if the track was farther from the fiber, the variation in photon-fiber distance remained relatively small. As a result, the Z readout showed stronger dependence on the passed position, leading to larger variation.
Turning to the Y readout, some samples exhibited large variance, while their X readout variation was similar to that of others.
Figure~\ref{fig:position_dependence} shows the position dependence of the mean light yield in the Y readout of samples T2 and T4.
Ideally, if the position dependence of the light yield arose solely from the distance to the fiber, the light yield in the Y readout would depend on the X position as observed for sample T4 in Fig.~\ref{fig:position_dependence_b}. 
However, the light yield of sample T2 in Fig.~\ref{fig:position_dependence_a}, which showed large variance in the Y direction, also depended on the Y position.
This result suggests that the WbLS state becomes unstable and the liquid scintillator concentration differs from top to bottom of the cell in these samples.
Other samples that had large variances in the Y readout, such as S1, T2 and T3, had the same tendency.

\subsubsection{Optical cross-talk between cells}
\begin{figure}[htbp]
\begin{center}
    \begin{minipage}[t]{0.35\linewidth}
    \begin{center}
        \includegraphics[width=0.9\linewidth]{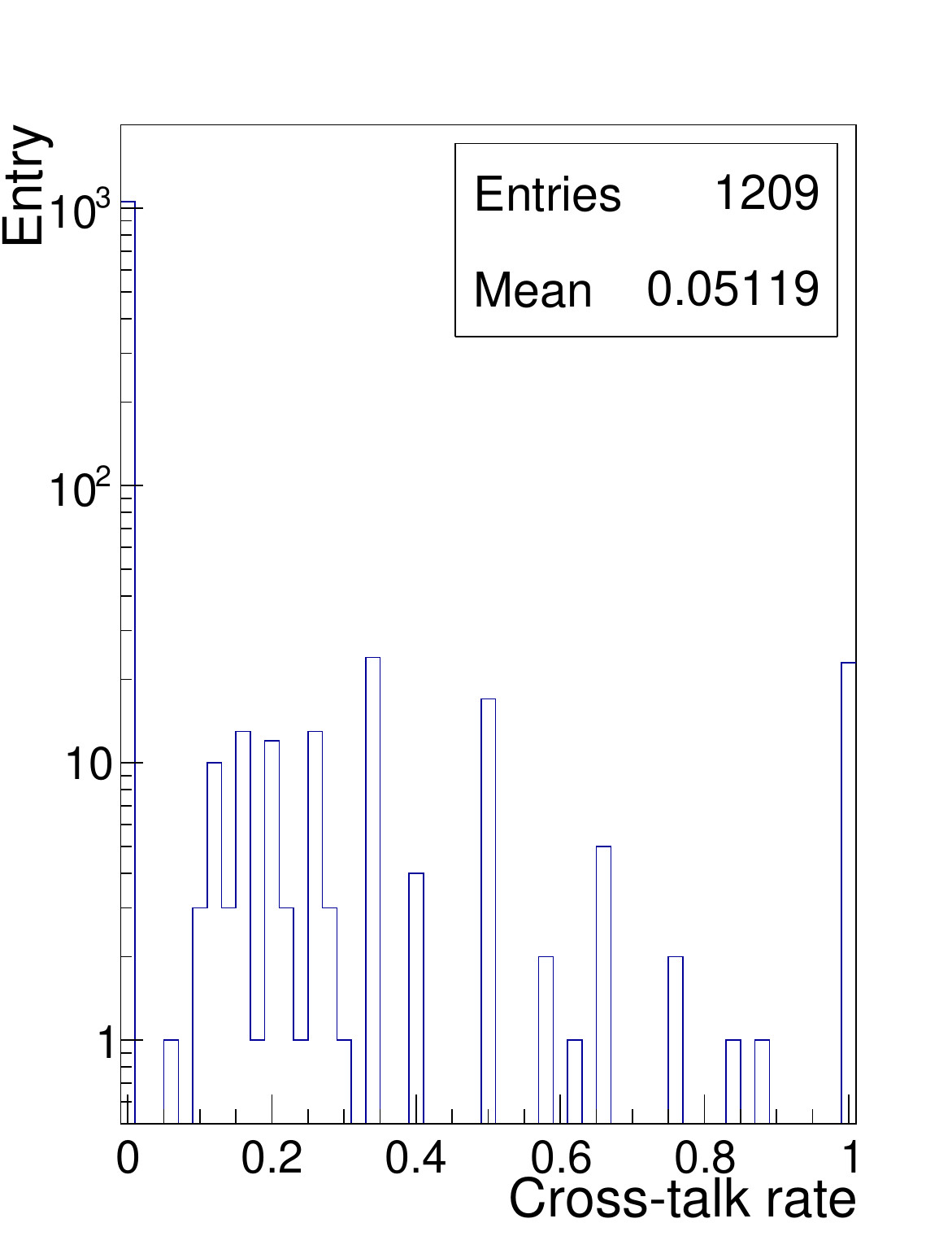}
        \vspace{5pt}
        \caption{Distribution of cross-talk rate between cells. }
        \label{fig:cross-talk_ex}
    \end{center}
    \end{minipage}
    \hspace{10pt}
    \begin{minipage}[t]{0.6\linewidth}
    \begin{center}
        \begin{minipage}[b]{0.38\linewidth}
            \includegraphics[width=0.9\linewidth]{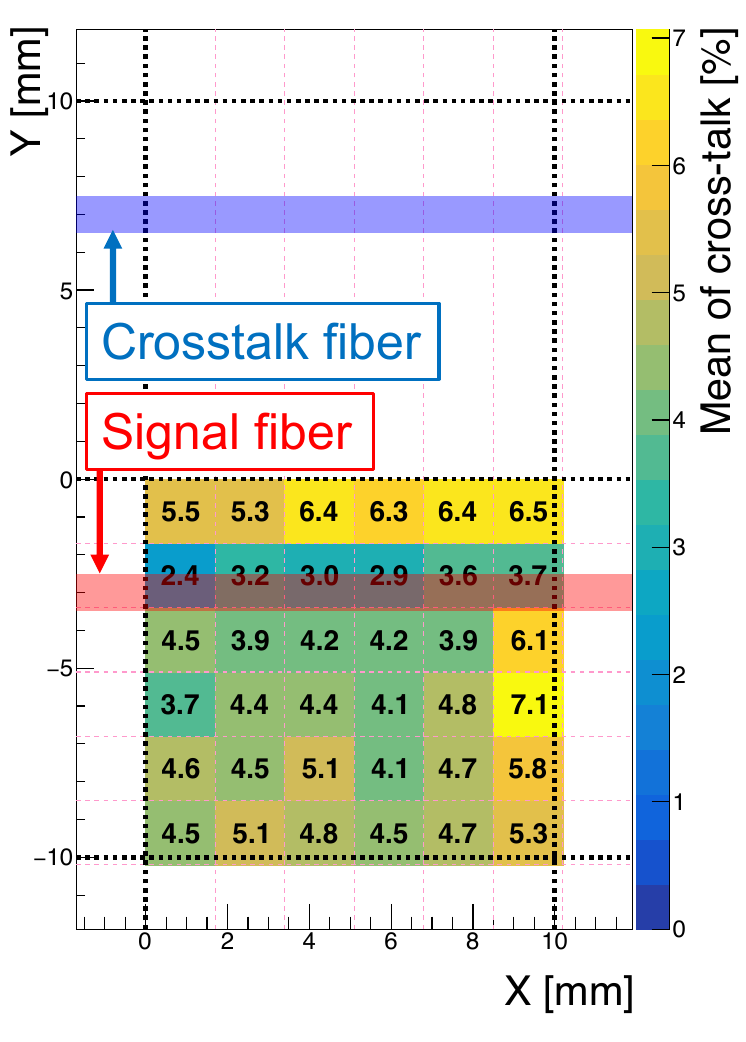}
            \subcaption{X readout}
        \end{minipage}
        \begin{minipage}[b]{0.6 \linewidth}
            \includegraphics[width=0.9\linewidth]{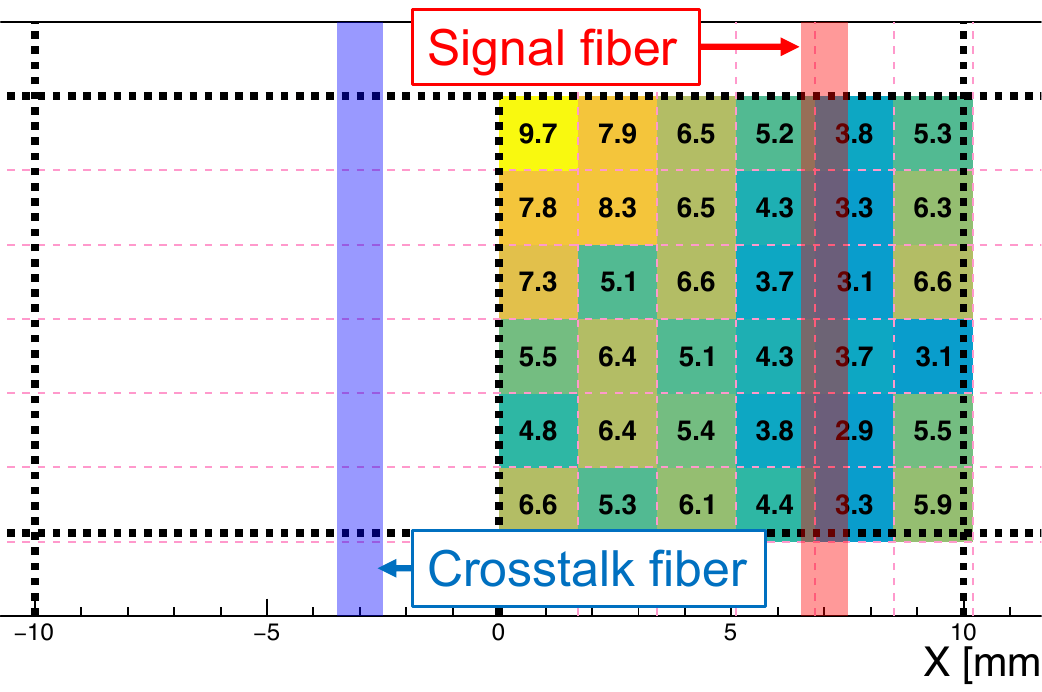}
            \subcaption{Y readout}
            \includegraphics[width=0.9\linewidth]{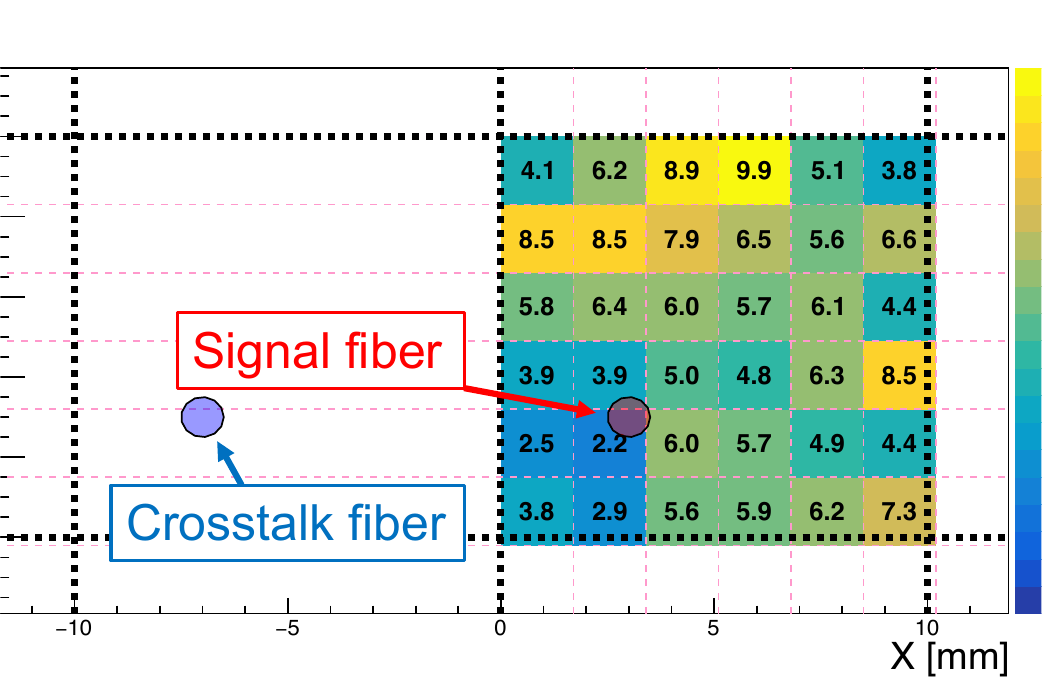}
            \subcaption{Z readout}
        \end{minipage}
        \vspace{-10pt}
        \hspace{10pt}\\
        \caption{Position dependence of the cross-talk rate between cells. }
        \label{fig:cross-talk}
    \end{center}
    \end{minipage}
\end{center}
\end{figure}

The role of the optical separator in the WbLS tracking detector is to trap photons in the cell.
However, with some probability, photon leakage into neighboring cells happens. 
Optical cross-talk rate between adjacent cells was evaluated by calculating the ratio of light yield in neighboring cells to that in the primary hit cell.
As shown in Fig.~\ref{fig:cross-talk_ex}, most events showed no signal in neighboring cells. 
However, occasional high cross-talk rates larger than 1 were observed, likely due to SiPM dark current or delta rays.
To avoid values being affected by these large value, the mean cross-talk rate was calculated using events with a cross-talk rate 1 or less.
Figure~\ref{fig:cross-talk} shows the position dependence of the mean cross-talk rate.
The cross-talk often happens near the fiber holes and the neighboring cells.
The mean cross-talk probability of the prototype WbLS detector is roughly 5\% and less than 10\%.
This is larger than 3\% cross-talk reported for the plastic scintillator cube of SuperFGD~\cite{bib:sfg_opt_sim}, indicating limited optical isolation with the current 3D-printed acrylic optical separator.

\section{Further improvement in the light yield of the WbLS tracking detector} \label{sec_improve}

The positron beam test revealed that the light yield from the prototype WbLS tracking detector was approximately one-third of the requirement for reliable single-cell detection of MIPs.
The low light yield was attributed to two main factors: (1)
insufficient total scintillation photon production and (2) low photon collection efficiency within the cell.
To address these issues, we developed new WbLS samples with higher light yield and evaluated potential improvements by using high-reflectivity materials as the optical separator.

\subsection{WbLS with high liquid scintillator concentration} \label{sec_improve_wbls}
The original WbLS samples used in the beam test consist of 70\% water and 20\% surfactant, and 10\% liquid scintillator (Fig.~\ref{fig:WbLS_beamtest}). 
To improve performance, we decreased the surfactant ratio and increased the liquid scintillator content.
To achieve this, we tested various surfactants including the ones not made in Sec.~\ref{sec_development} and found that the surfactant IGEPAL CO-630 has a higher solubility compared to the Triton X-100, the surfactant used in the beam test.

\begin{figure}[htbp]
  \begin{center}
    \begin{minipage}{0.35\linewidth}
        \begin{center}
            \includegraphics[width=0.95\linewidth]{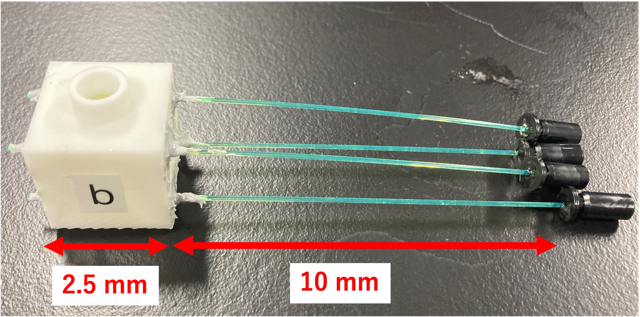}
            \caption{A cell used for the light yield measurement of WbLS samples with a high liquid scintillator ratio.}
            \label{fig:improved_setup_cell}
        \end{center}
    \end{minipage}
    \begin{minipage}[c]{0.63\linewidth}
        \begin{center}
            \includegraphics[width=0.95\linewidth]{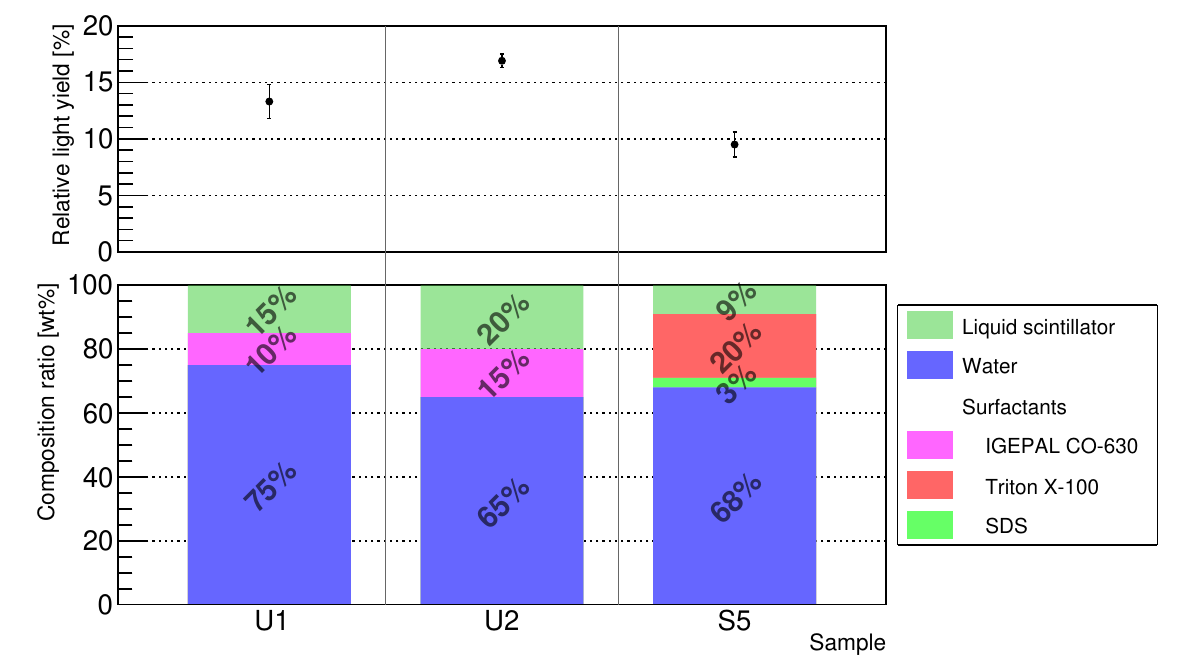}
            \caption{Composition of WbLS and the measured light yield relative to the pure liquid scintillator. }
            \label{fig:improved_wbls}
        \end{center}
    \end{minipage}
  \end{center}
\end{figure}

The light yield of newly formulated WbLS samples with high liquid scintillator concentration was tested using cosmic rays with the WLS fiber readout system.
The test cell, shown in Fig.~\ref{fig:improved_setup_cell}, was larger than the 1~cm$^3$ cells and included four fibers to facilitate light yield comparison. The cell was placed vertically between two trigger plastic scintillators.

Figure~\ref{fig:improved_wbls} shows the compositions and the light yield of new U1 and U2 samples, as well as for the S5 sample, which had the same composition as S1 sample used in the beam test. 
The U2 sample which has 65\% water and 20\% liquid scintillator had a $1.78\pm0.22$ times higher light yield than the S5 sample while losing 3\% of water ratio. 

\subsection{High reflectivity optical separator} \label{sec_improve_reflectivity}
Scintillation photons are usually reflected many times by the optical separator before being captured by the WLS fiber, since photons are emitted isotropically and the viewing angle toward the fiber is small. 
Therefore, the reflectivity of the optical separator can significantly affect collection efficiency inside cell.

We evaluated this effect through Geant4-based optical simulations using the same two- and three-directional readouts cells used as in the beam test.
In this simulation, the optical absorption length was set to that of pure water (more than 10~m at the wavelength of scintillation photons), and the scattering length was assumed to be infinite. 
The simulation showed that the impact on light yield was negligible as long as the absorption length exceeded 1~m and scattering length exceeded 1~cm.

\begin{figure}[htbp]
    \begin{center}
        \includegraphics[width=0.6\linewidth]{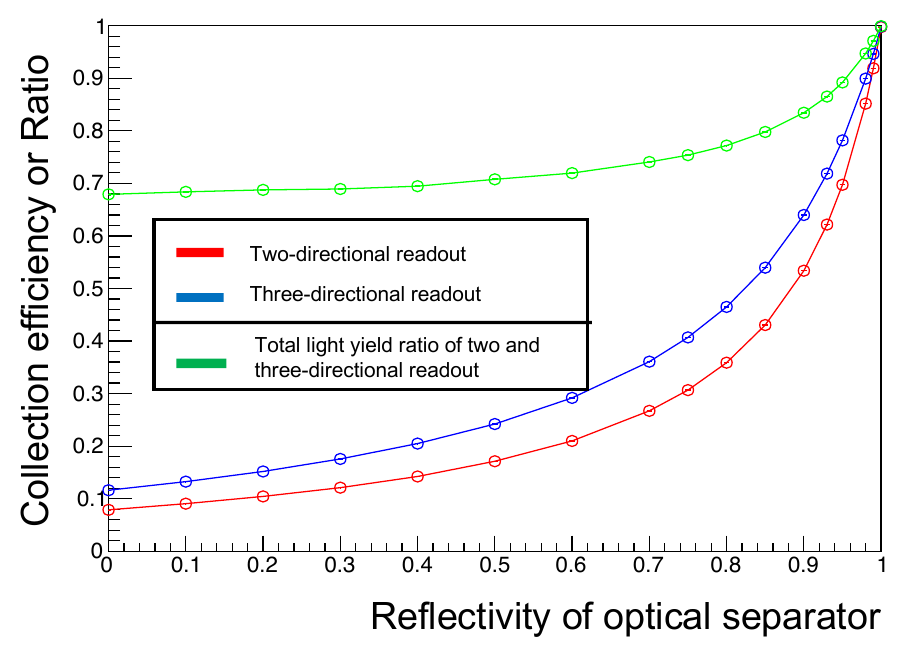}
         \caption{
            Simulation result of the collection efficiency of scintillation photons produced in the cell as a function of the reflectivity of the optical separator. 
         }
         \label{fig:optical_separator_reflectivity}
    \end{center}
\end{figure}

As shown in Fig.~\ref{fig:optical_separator_reflectivity}, the collection efficiency increased 
significantly when the reflectivity approached unity. 
As noted in Sec.~\ref{sec_beamtest_ly}, the comparison between the light yields of two- and three-directional readouts implied a low collection efficiency, likely due to the low reflectivity of the optical separator. 
Therefore, there is considerable potential for improving the light yield of the WbLS tracking detector by using a high-reflectivity material as the optical separator.

To demonstrate this, we measured the light yield using a cell equipped with PTFE reflective sheet which have 93\% reflectivity.
The measured light yield was approximately 1.5 times higher than that obtained with the cell made by the same material in the beam test.
However, it is impractical to construct the WbLS tracking detector with this reflective sheet, so further investigation into 3D-printable reflective materials or a method to apply reflective coatings inside the cell is needed. 

\subsection{Projected performance and outlook} \label{sec_summary_outlook}

Based on the results presented in Sec.~\ref{sec_improve_wbls} and Sec.~\ref{sec_improve_reflectivity}, the total light yield of the WbLS tracking detector is expected to increase by a factor of approximately 2.5, corresponding to about 6.0~p.e./MeV per fiber readout. 
This projection reflects the combined improvements in scintillation photon production and photon collection efficiency, achieved through enhanced WbLS composition and increased separator reflectivity.

However, this yield remains below the estimated requirement of 8.1~p.e./MeV per fiber readout necessary to achieve 99\% detection efficiency for MIPs.
Addressing this gap could be achieved with continued improvement across detectors.

\section{Conclusions}\label{sec_conclusion}
We are developing a new tracking detector using WbLS with a WLS fiber readout for a precise measurement of neutrino interactions on a water target.
We produced a variety of WbLS samples with different material compositions and measured their light yield using both PMT and fiber readouts.
A prototype WbLS tracking detector consisting of 4$\times$4$\times$5 cells was fabricated and its performance was evaluated using a positron beam at RARiS.
We successfully observed particle tracks for the first time using WbLS with a fiber readout.
However, the measured light yield was as low as 2.4 p.e./MeV per fiber readout, while 8.1 p.e./MeV is required to detect MIPs with 99\% efficiency.
In addition, the optical cross-talk rate was around 5\%.
To address these issues, we improved both the WbLS composition and the optical separator, achieving light yield improvements of $1.78\pm0.22$ and 1.5 times, respectively.
Although the achievable light yield of 6.0 p.e./MeV is still below the requirement, these results reported here indicate steady progress toward the performance necessary for a viable WbLS tracking detector for future neutrino experiments.

\section*{Acknowledgment}
Part of this study was performed using facilities of Research Center for Accelerator and Radioisotope Science (RARiS), Tohoku University (Proposal No. 3006).
The authors are grateful to the RARiS staffs for supplying a superb beam.
In addition, the authors would like to thank Davide Sgalaberna for valuable discussions throughout this study.
This work was supported by MEXT KAKENHI Grant Number JP22K14058, JP24K00644 and Kyoto University's internal fund, ISHIZUE FY2021.


\begin{thebibliography}{99}


  \bibitem{bib:yeh} M.~Yeh~\etal, Nucl. Instrum. Meth. A 660, 51 (2011)
  \bibitem{bib:hans} H. Th. J. Steiger~\etal, arXiv:2405.05743 (2024). 
  
  \bibitem{bib:hk} K.~Abe~\etal~(Hyper-Kamiokande Proto-Collaboration), Hyper-Kamiokande Design Report, arXiv:1805.04163 (2018)

  \bibitem{bib:t2k} K.~Abe~\etal~(T2K Collaboration), Nucl. Instrum. Meth. A {\bf 659}  106 (2011).
  
  \bibitem{bib:tdr} K.~Abe~\etal, arXiv:1901.03750 (2019). 
  \bibitem{bib:sfgd} A.~Blondel~\etal, JINST {\bf 13} P02006 (2018).
  \bibitem{bib:sfgd_kikawa} T.~Kikawa~\etal, Nucl. Instrum. Meth. A {\bf 1080}, 170616 (2025).  
  \bibitem{bib:easiroc} I.~Nakamura~\etal, Nucl. Instrum. Meth. A {\bf 787}, 376 (2015).
  \bibitem{bib:sfg_opt_sim} S.~Abe~\etal, Nucl. Instrum. Meth. A {\bf 1080}, 170757 (2025).
  \bibitem{geant4_1} S.~Agostinelli~\etal, Nucl. Instrum. Meth. A 506 (2003) 250-303.
  \bibitem{geant4_2} J.~Allison~\etal, IEEE Trans. Nucl. Sci. 53 (2006) 270-278.
  \bibitem{geant4_3} J.~Allison~\etal, Nucl. Instrum. Meth. A 835 (2016) 186-225.

\end{thebibliography}
\end{document}